\documentclass{aa}

\usepackage[utf8]{inputenc}
\usepackage{graphicx}
\usepackage{appendix}
\usepackage{booktabs}
\usepackage{txfonts}
\usepackage{threeparttable}
\usepackage{longtable}
\usepackage{tabularx}
\usepackage{multirow}
\usepackage{multicol}
\usepackage{comment}
\usepackage{xcolor}
\usepackage{placeins}
\usepackage{float}

\begin{document}

   \title{First extragalactic detection of a phosphorus-bearing molecule with ALCHEMI: Phosphorus nitride (PN)}

   \author{D. Haasler\inst{\ref{inst.URJC},\ref{inst.MadridCAB}}
        \and V. M. Rivilla\inst{\ref{inst.MadridCAB},\ref{inst.INAF}}
        \and S. Mart\'in\inst{\ref{inst.ESOChile},\ref{inst.JAO}}
         \and J. Holdship\inst{\ref{inst.Leiden},\ref{inst.UCL}}
        \and S. Viti\inst{\ref{inst.Leiden},\ref{inst.UCL}}
        %---------- PIs
        \and N. Harada \inst{\ref{inst.NAOJ},\ref{inst.ASIAA},\ref{inst.SOKENDAI}}
        \and J. Mangum \inst{\ref{inst.NRAOCV}}
        %----------DATA REDUCTION/PIs
        \and K.~Sakamoto \inst{\ref{inst.ASIAA}}
        \and S.~Muller \inst{\ref{inst.ONSALA}}
        \and K.~Tanaka \inst{\ref{inst.KeioUniversity}}
        \and Y. Yoshimura  \inst{\ref{inst.UTokio}}
        \and K. Nakanishi \inst{\ref{inst.NAOJ},\ref{inst.SOKENDAI}}
%        \and ... 
     %----------
        \and L. Colzi \inst{\ref{inst.MadridCAB},\ref{inst.INAF}}
        \and L. Hunt \inst{\ref{inst.INAF}}
        \and K.~L.~Emig\inst{\ref{inst.NRAOCV}}\thanks{Jansky Fellow of the National Radio Astronomy Observatory}
        \and R.~Aladro \inst{\ref{inst.MaxPlanck}}
        \and P.~Humire \inst{\ref{inst.MaxPlanck}}
        \and C.~Henkel \inst{\ref{inst.MaxPlanck},\ref{inst.Abdulaziz},\ref{inst.XinjiangObs}}
        \and P.~van der Werf \inst{\ref{inst.Leiden}}
        }
\institute{
\label{inst.URJC}Escuela Superior de Ciencias Experimentales y Tecnología (ESCET), Universidad Rey Juan Carlos, Tulipán s/n, 28933 Móstoles, Madrid, Spain 
\and\label{inst.MadridCAB}Centro de Astrobiolog\'ia (CSIC-INTA), Ctra. de Ajalvir Km. 4, Torrej\'on de Ardoz, 28850 Madrid, Spain
\and\label{inst.INAF}INAF-Osservatorio Astrofisico di Arcetri, Largo Enrico Fermi 5, 50125, Florence, Italy
\and\label{inst.ESOChile}European Southern Observatory, Alonso de C\'ordova, 3107, Vitacura, Santiago 763-0355, Chile
\and\label{inst.JAO}Joint ALMA Observatory, Alonso de C\'ordova, 3107, Vitacura, Santiago 763-0355, Chile
\and\label{inst.Leiden}Leiden Observatory, Leiden University, PO Box 9513, NL - 2300 RA Leiden, The Netherlands
\and\label{inst.UCL}Department of Physics and Astronomy, University College London, Gower Street, London WC1E6BT, UK
\and\label{inst.NAOJ}National Astronomical Observatory of Japan, 2-21-1 Osawa, Mitaka, Tokyo 181-8588, Japan
\and\label{inst.ASIAA}Institute of Astronomy and Astrophysics, Academia Sinica, 11F of AS/NTU Astronomy-Mathematics Building, No.1, Sec. 4, Roosevelt Rd, Taipei 10617, Taiwan   
\and\label{inst.SOKENDAI}Department of Astronomy, School of Science, The Graduate University for Advanced Studies (SOKENDAI), 2-21-1 Osawa, Mitaka, Tokyo, 181-1855 Japan
\and\label{inst.NRAOCV}National Radio Astronomy Observatory, 520 Edgemont Road, Charlottesville, VA 22903-2475, USA
\and\label{inst.ONSALA}Department of Space, Earth and Environment, Chalmers University of Technology, Onsala Space Observatory, SE-43992 Onsala, Sweden
\and\label{inst.KeioUniversity}Department of Physics, Faculty of Science and Technology, Keio University, 3-14-1 Hiyoshi, Yokohama, Kanagawa 223--8522 Japan
\and\label{inst.UTokio}Institute of Astronomy, Graduate School of Science, The University of Tokyo, 2-21-1 Osawa, Mitaka, Tokyo 181-0015, Japan
\and\label{inst.MaxPlanck}Max-Planck-Institut f\"{u}r Radioastronomie, Auf dem H\"{u}gel 69, D-53121 Bonn, Germany
\and\label{inst.Abdulaziz}Astron. Dept., Faculty of Science, King Abdulaziz University, P.O. Box 80203, Jeddah 21589, Saudi Arabia
\and\label{inst.XinjiangObs}Xinjiang Astronomical Observatory, Chinese Academy of Sciences, Urumqi 830011, P.R. China}

   \date{Received; accepted}

% \abstract{}{}{}{}{} 
% 5 {} token are mandatory
 
   \abstract
   {Phosphorus (P) is a crucial element for life given its central role in several biomolecules. P-bearing molecules have been discovered in different regions of the Milky Way, but not yet towards an extragalactic environment.}
   {We searched for P-bearing molecules outside the Milky Way towards the nearby starburst Galaxy NGC 253.}
   {Using observations from the ALMA Comprehensive High-resolution Extragalactic Molecular Inventory (ALCHEMI) project, we used the MAdrid Data CUBe Analysis (MADCUBA) package to model the emission of P-bearing molecules assuming local thermodynamic equilibrium (LTE) conditions. We also performed a non-LTE analysis using SpectralRadex.}
   {We report the detection of a P-bearing molecule, phosphorus nitride (PN), for the first time in an extragalactic environment, towards two giant molecular clouds (GMCs) of NGC 253. The LTE analysis yields total PN beam-averaged column densities $N$=(1.20$\pm$0.09)$\times$10$^{13}$ cm$^{-2}$ and $N$=(6.5$\pm$1.6)$\times$10$^{12}$ cm$^{-2}$, which translate into abundances with respect to H$_2$ of $\chi$=(8.0$\pm$1.0)$\times$10$^{-12}$ and $\chi$=(4.4$\pm$1.2)$\times$10$^{-12}$. We derived a low excitation temperature of $T_{\rm ex}$=(4.4$\pm$1.3) K towards the GMC with the brightest PN emission, which indicates that PN is sub-thermally excited. The non-LTE analysis results in column densities consistent with the LTE values. We also searched for other P-bearing molecules (PO, PH$_{3}$, CP, and CCP), and upper limits were derived. The derived PO/PN ratios are $<$1.3 and $<$1.7.
   The abundance ratio between PN and the shock-tracer SiO derived towards NGC 253 follows the same trend previously found towards Galactic sources. 
   }
   {Comparison of the observations with chemical models indicates that the derived molecular abundances of PN in NGC 253 can be explained by shock-driven chemistry followed by cosmic-ray-driven photochemistry.
   }


   \maketitle
%
%-------------------------------------------------------------------

\section{Introduction}

Phosphorus (P), along with oxygen, hydrogen, carbon, nitrogen, and sulfur, is one of the six biogenic elements and it is essential for the development of life.
It has a pivotal role in key biomolecules such as phospholipids that conform the structure of cellular membranes, nucleic acids (e.g. DNA and RNA), or adenosine triphosphate (ATP), which is the molecular energy currency in cells \citep{bookMacia2019}.
Although P is not abundant in the Universe, it has a vital importance in biology on Earth.
Its cosmic abundance compared to hydrogen (H) is $\sim$2.6$\times$10$^{-7}$ \citep{Asplund2009}, being only the nineteenth most abundant element. For this reason, the detection of P-bearing molecules in the interstellar medium (ISM) is challenging, and as a consequence its interstellar chemistry is still not fully understood. While it is known that P is synthesised in the interior of massive stars (\citealt{Koo2013}) and injected into the ISM through supernova explosions, it is still a matter of debate how it is incorporated into molecules in molecular clouds (e.g. \citealt{Rivilla2020}).

To date, several P-bearing molecules have been found in the ISM of our Galaxy, starting with the first simultaneous phosphorus nitride (PN) detections towards massive star-forming regions simultaneously by \cite{Turner1987} and \cite{Ziurys1987}. Afterwards, PN was also detected towards protostellar shocks (\citealt{Yamaguchi2011,Lefloch2016}), Galactic Centre molecular clouds (\citealt{Rivilla2018}), envelopes of evolved stars \citep{Milam2008,Ziurys2018}, multiple massive star-forming regions (\citealt{Fontani2016,Mininni2018,Fontani2019,Rivilla2020,Bernal2021}), and the surrounding environment of low-mass protostars (\citealt{Bergner2019}).

The first detection of interstellar phosphorus oxide (PO) was reported by \cite{Tenenbaum2007} in the circumstellar region of the evolved star VY Canis Majoris. Later, PO was also detected in several massive star-forming regions (\citealt{Rivilla2016,Rivilla2020,Bernal2021}), a protostellar shock (\citealt{Lefloch2016}), a Galactic Centre molecular cloud (\citealt{Rivilla2018}), and a low-mass star-forming region (\citealt{Bergner2019}). Moreover, PO has also been detected in the coma of the comet Churyumov-Gerasimenko recently thanks to in situ measurements by the Rosetta spacecraft (\citealt{Rivilla2020}).
In all these different environments, it has been found that the molecular abundance PO/PN ratio is always higher than 1.

Other P-bearing molecules have also been detected in our Galaxy, but only in the circumstellar envelopes of evolved stars and/or planetary atmospheres. Phosphine (PH$_{3}$) was detected towards IRC+10216 \citep{Agundez2014} and in the atmosphere of the gaseous planets Saturn \citep{Bregman1975} and Jupiter \citep{Larson1977}. In addition CP, CCP, and HCP have also been detected in the envelope of evolved stars \citep{Agundez2007,Agundez2014,Halfen2008,Milam2008}.

So far, all of the detections of P-bearing molecules have been reported in regions of our Galaxy. Although atomic P has been detected towards extragalactic environments, for example in the Large Magellanic Cloud (LMC) (\citealt{Friedman2000}) and in a damped-Lyman absorber towards the quasar QSO 0000-2620 (\citealt{Molaro2001}), P-bearing molecules outside our Galaxy still have yet to be found. The improved sensitivity of current radiotelescopes (e.g. the Atacama Large Millimeter/sub-millimeter Array, ALMA) allows us to search for P-bearing species beyond our Galaxy. We present in this work a search for P-bearing molecules towards the galaxy NGC~253 using the spectral line survey of the ALMA Comprehensive High-resolution Extragalactic Molecular Inventory (ALCHEMI; \citealt{Martin2021}).
NGC~253 is a nearly edge-on barred spiral galaxy and a prototype of nearby starburst \citep[distance of 3.5 Mpc;][]{Rekola2005}. Its star formation rate of $2~\rm M_\odot~yr^{-1}$ within its central kiloparsec is roughly twice that within the entire Milky Way \citep{Robitaille2010,Leroy2015,Bendo2015}.  
The metallicity of NGC~253 is consistent with or just above Solar \citep[$\rm 12+log(O/H)=8.99\pm0.31$;][]{Marble2010}.
The central molecular zone (CMZ) of NGC 253 is one of the most prolific extragalactic sources for detections of molecules, as shown by previous spectral surveys, which revealed a rich chemical composition (e.g. \citealt{Martin2006,Aladro2015,Meier2015,Ando2017,Martin2021}). This fact makes NGC~253 an excellent target for the search of new molecular species in the extragalactic ISM.

\section{Observations}

We used observations from the ALMA Comprehensive High-resolution Extragalactic Molecular Inventory (ALCHEMI), which is an ALMA Large Program to study the molecular inventory of the starburst Galaxy NGC 253 (\citealt{Martin2021}). The observations target the CMZ of NGC 253 covering a rectangular area of $50'' \times 20''$, that is 850 $\times$ 340 pc centred at RA(ICRS) = 00$^h$47$^m$33.26$^s$, DEC(ICRS) = $-$25º17$'$17.7$''$.
ALCHEMI is an unbiased spectral survey that uses the ALMA main array and ACA (ALMA Compact Array) configurations to fully cover the bands 3, 4, 5, 6, and 7 (from 84.2 to 373.2 GHz) down to an average sensitivity of $\sim$15~mK in 10~km~s$^{-1}$ channels at a common angular resolution of $1.6''$ (28 pc). More details about the observations and the data reduction of ALCHEMI are extensively presented in \cite{Martin2021}.

\section{Data analysis and results}
\label{analysis}

We used MADCUBA\footnote{MADCUBA (Madrid Data Cube Analysis) is a software developed in Centro de Astrobiología (INTA-CSIC, Madrid) to visualise and analyse single spectra and datacubes.\\ \url{https://cab.inta-csic.es/madcuba/}} \citep{Martin2019b}
for the inspection of the ALCHEMI datacubes, the identification of the molecular species, and the LTE derivation of the physical parameters of the emission.
The Spectral Line Identification and Modelling (SLIM) tool of MADCUBA generates a simulated spectrum assuming local thermodynamic equilibrium (LTE) conditions. The AUTOFIT tool performs a non-linear least-squares fitting of the simulated LTE spectrum to the observed data utilising the Levenberg-Marquardt algorithm \citep{Levenberg1944,Marquardt1963} and provides the best solution for the model parameters: total molecular column density (\textit{N}), excitation temperature (\textit{T$_{\rm ex}$}), line velocity (v$_{\rm LSR}$), line full width at half maximum (FWHM), and source size ($\theta_s$).
Table \ref{table:spectroparam} lists the different molecular transitions analysed in this work, their frequencies, upper energy levels, and Einstein $A_{\rm ul}$ coefficients.

\begin{table}
\begin{center}
\begin{threeparttable}
\caption{Molecular transitions used in the analysis.}
\begin{tabular}{ c c c c c } 
\hline
\multirow{2}{4em}{\centering Molecule} & \multirow{2}{4em}{\centering Transition} & Frequency & E$_{\rm up}$ & $A_{\rm ul}$ \\
 &  & (GHz) & (K) & ($\times$10$^{-5}$ s$^{-1}$) \\ %[2pt]
  \hline
%  \noalign{\vskip 4pt}
  \multirow{2}{4em}{\centering PN} & J=2$-$1 & 93.979 & 6.8 & 2.9 \\
  & J=3$-$2 & 140.967 & 13.5 & 10.5 \\
  \noalign{\vskip 4pt}
  \multirow{2}{4em}{\centering PO$^{a}$} & F=3$-$2, l=e & 108.998 & 8.4 & 2.1 \\
  & F=2$-$1, l=e & 109.045 & 8.4 & 1.9\\
  \noalign{\vskip 4pt}
  \centering PH$_{3}$ & J=1$-$0 & 266.945 & 12.8 & 2.4 \\
  \noalign{\vskip 4pt}
  \multirow{2}{4em}{\centering CP$^{b}$} & F=3$-$2 & 95.710 & 6.9 & 0.4 \\
  & F=2$-$1 & 95.713 & 6.9 & 0.4 \\
  \noalign{\vskip 4pt}
  \multirow{4}{4em}{\centering CCP$^{c}$} & F=8$-$7, l=e & 95.406 & 19.2 & 5.3 \\
  & F=7$-$6, l=e & 95.410 & 19.3 & 5.2 \\
  & F=8$-$7, l=f & 95.456 & 19.3 & 5.3 \\
  & F=7$-$6, l=f & 95.460 & 19.3 & 5.3 \\
  \noalign{\vskip 4pt}
  \multirow{3}{4em}{\centering $^{29}$SiO} & J=2$-$1 & 85.759 & 6.2 & 2.8 \\
   & J=3$-$2 & 128.637 & 12.3 & 10.2 \\
   & J=4$-$3 & 171.512 & 20.6 & 25.1 \\
  \noalign{\vskip 4pt}
  \multirow{2}{4em}{\centering C$^{34}$S} & J=2$-$1 & 96.412 & 6.2 & 1.6 \\
   & J=3$-$2 & 144.617 & 11.8 & 5.8 \\
%  \noalign{\vskip 3pt}
  \hline
%  \noalign{\vskip 5pt}
\end{tabular}
$^{a}$ J=5/2$-$3/2, $\Omega$=1/2. The other two lines of this quadruplet at 109.206 and 109.281 GHz are not used due to heavy contamination from other molecular species.\\
$^{b}$ J=5/2$-$3/2.\\
$^{c}$ J=15/2$-$13/2. $\Omega$=1/2 \\
Spectroscopic data obtained from the CDMS catalogue: PN entry 045511 (September 2019), PO entry 047507 (October 2019), PH$_{3}$ entry 034501 (May 2013), CP entry 043501 (March 2000), CCP entry 055503 (April 2009), $^{29}$SiO entry 045504 (January 2014), and C$^{34}$S entry 046501 (January 2004).
\label{table:spectroparam}
\end{threeparttable}
\end{center}
\end{table}

\subsection{Detections of phosphorus nitride (PN) and LTE analysis}
\label{sec:analysis-LTE}

\begin{figure*}
\centering
\includegraphics[width=\textwidth]{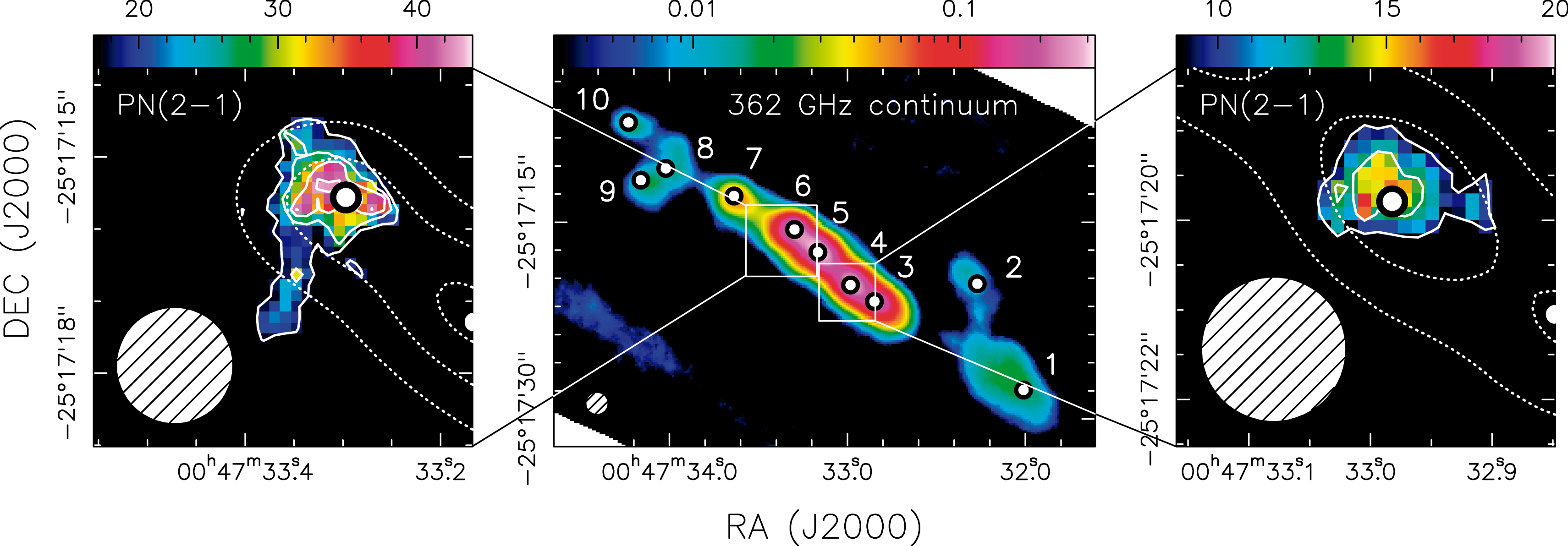}
  \caption{{\it Centre:} Continuum map of the CMZ of NGC 253 at 362 GHz with a coloured scale indicating the emission in Jy beam$^{-1}$.
  The numbered circles represent the brightest points of each GMC detected by \cite{Leroy2015}. The synthesised beam of the observations (1.6$''$) is shown in the lower left corner of each panel with a white hatched circle.
  {\it Left:} Zoomed in view towards GMC 6 of the integrated emission between 120-240 km s$^{-1}$. {\it Right:} Zoomed in view towards GMC 4 of the integrated emission between 220-260 km s$^{-1}$. Figures on the sides have their integrated emission scale in mJy beam$^{-1}$ km s$^{-1}$ starting at 2$\sigma$. The white dashed contour levels indicate the continuum emission for 0.1, 0.2, and 0.3 Jy beam$^{-1}$.
  Solid white contours indicate 2$\sigma$, 3$\sigma$, 4$\sigma$ and 5$\sigma$ levels of the PN detection.}
     \label{figure:map}
\end{figure*}

Using the datacube visualisation tool of MADCUBA, we inspected the CMZ of NGC 253  to search for PN emission towards the Giant Molecular Clouds (GMCs) shown in Fig. \ref{figure:map} (we have adopted the labelling from \citealt{Leroy2015}), with the exception of GMC 5.
This region is located at the position of the dynamical centre of NGC~253 determined by \citealt{MullerSanchez2010}. This region contains several bright (thermal) radio continuum sources that are parsecs in size \citep{Ulvestad1997, Mills2021}; when this region is observed at $\sim$1$''$ resolution, molecular components of the CMZ appear in absorption \citep{Mangum2019}. Because this results in complex spectral line profiles from a variety of molecular species, we do not analyse GMC 5.
We used the PN spectroscopic data provided by the Cologne Database for Molecular Spectroscopy (CDMS) catalogue\footnote{\url{https://cdms.astro.uni-koeln.de/classic/catalog}} \citep{Muller2001,Endres2016} entry 045511 (version of September 2019) based on the works by \cite{Cazzoli2006} and the dipole moment from \cite{Hoeft1972}.  

We successfully detected PN towards GMCs 4 and 6. The side panels in Fig. \ref{figure:map} present the integrated emission maps of PN(2-1) towards these two molecular clouds. The PN(2$-$1) transition is not contaminated by other species, as seen in Galactic star-forming regions (e.g. \citealt{Fontani2016,Fontani2019,Rivilla2020}) and in Galactic Centre molecular clouds \citep{Rivilla2018}. The emission is enhanced towards the continuum peaks in both GMCs. In GMC 6 it exhibits a tail-like structure towards the south-east at a 2$\sigma$ level. This emission might be a real feature, but the current sensitivity does not allow us to draw a firm conclusion.

\begin{figure}
    \centering
    \includegraphics[width=8.8cm]{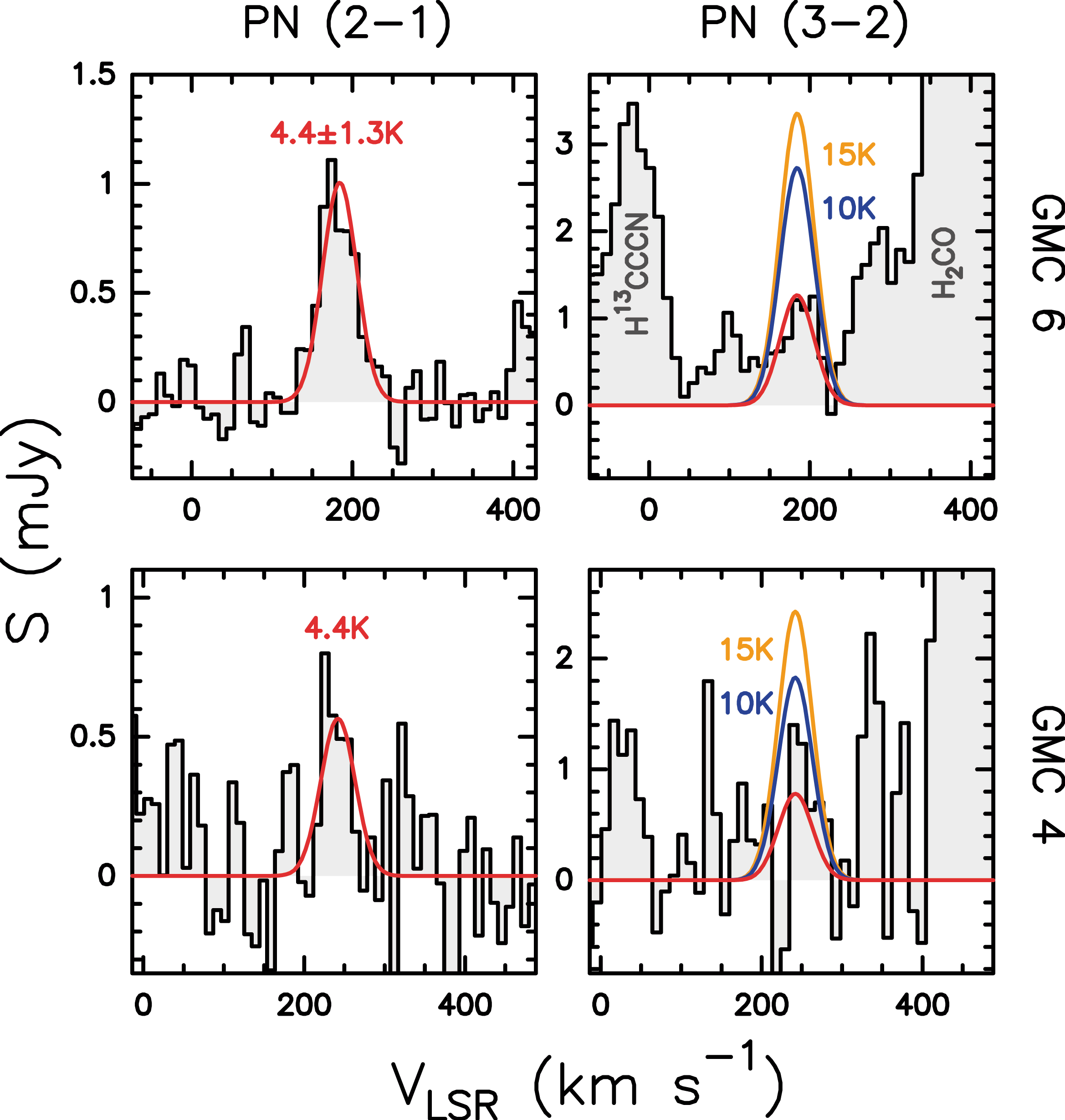}
      \caption{PN(2-1) (left) and PN(3-2) (right) transitions towards the GMCs 6 (upper panels) and 4 (lower panels) of NGC 253. The spectra have been extracted towards the peak of the continuum.
      The best LTE fits to the line emission are indicated in red. In blue and yellow we show the expected line profiles for PN(3-2) assuming $T_{\rm ex}$ of 10 K and 15 K, respectively.}
    \label{figure:PN}
\end{figure}

To perform the  molecular analysis, we extracted the spectra towards the pixel of the peak of the continuum emission of the GMCs.
The spectra of the PN(2$-$1) transition towards GMCs 6 and 4 are displayed in the left panels of Fig. \ref{figure:PN}. 
The transition is detected at levels of 16$\sigma$ and 5$\sigma$ in integrated intensity for GMCs 6 and 4, respectively. 
The right panels of Fig. \ref{figure:PN} show the spectra of the PN(3$-$2) transition, which is detected in GMC 6 and not detected in GMC 4.

Transitions of PN up to the 7$-$6 fall also within the ALCHEMI spectral coverage, but their expected intensities, according to the fit obtained with the PN(2$-$1) and PN(3$-$2) transitions (see below), are $<$ 0.6 mJy, below the noise level of the data.
Moreover, some PN transitions appear heavily contaminated by transitions of more abundant species: PN(4$-$3) is contaminated by CH$_{3}$C$_{2}$H(11$-$10) and PN(6$-$5) by several transitions of HC$_{3}$N.

To derive the physical parameters of the emission, we used the MADCUBA-AUTOFIT tool. We fixed the source size to $\theta_s$=1.6$''$, which matches the ALCHEMI synthesised beam, and hence the derived molecular column densities are beam-averaged. 
We note that if the PN emission would be confined to the $0.6"$ core sizes measured at higher resolution \citep[clumps 1 and 6 in][]{Ando2017}, column densities in this work would be underestimated by a factor of $\sim$7. Since the uncertainty in the filling factor cancels when comparing species, assuming co-spatial distribution, the column density ratios between species still provide reliable constraints.
We also fixed the FWHM to the linewidth that reproduces well the PN(2$-$1) profiles, which is 50 km s$^{-1}$. For GMC 6, we left free the other parameters ($N$, $T_{\rm ex}$, and v$_{\rm LSR}$) while simultaneously fitting the PN(2$-$1) and PN(3$-$2) transitions. 
The resulting best LTE fit is shown with a red curve in the upper panels of Fig. \ref{figure:PN} and the derived parameters are listed in Table \ref{table:parameters}. We obtain a total PN column density \textit{N}=(1.20$\pm$0.09)$\times$10$^{13}$ cm$^{-2}$ and excitation temperature $T_{\rm ex}$=4.4$\pm$1.3 K.
Higher values of $T_{\rm ex}$ (e.g. 10 and 15 K) would result in higher intensities of the PN(3$-$2) transition, which are not consistent with the observations, as shown in Fig. \ref{figure:PN}.
The low $T_{\rm ex}$ derived is similar to the ones derived in Galactic Centre molecular clouds (\citealt{Rivilla2018}) and to high-mass and low-mass star forming regions (\citealt{Fontani2019,Bergner2019}).
Such a low $T_{\rm ex}$ suggests that the molecule is sub-thermally excited. 
We derived the relative abundance of PN compared to H$_{2}$, using the total column density of H$_{2}$, \textit{N$_{\rm H_2}$}, derived by \cite{Mangum2019} from dust continuum observations with a similar spatial resolution than ALCHEMI  (1.5$''\times$0.9$''$). Assuming that PN and dust trace similar volumes of gas (we discuss this assumption in Sect. \ref{sec:discussion}), we obtain an abundance of $\chi$=(0.8$\pm$0.1)$\times$10$^{-11}$  (see Table \ref{table:parameters}). 

For GMC 4, since the PN(3$-$2) is within the noise level (Fig. \ref{figure:PN}), we performed the LTE fit using only the PN(2$-$1) transition and fixing $T_{\rm ex}$ to the one derived in GMC 6. 
We obtain a total PN column density of \textit{N}=(6.5$\pm$1.6)$\times$10$^{12}$ cm$^{-2}$, which translates into a PN abundance of  $\chi$=(0.44$\pm$0.12)$\times$10$^{-11}$. The results of the fit for GMC 4 are presented in Table \ref{table:parameters}.

PN is not detected towards the rest of the GMCs, and therefore we provide upper limits for its column density in these regions. 
MADCUBA calculates the upper limit of the column density using the 3$\sigma$ value of the integrated intensity (see details in \citealt{Martin2019b}).
We fixed $T_{\rm ex}$ and FWHM to the same values as the ones derived in the PN detection of GMC 6, and v$_{\rm LSR}$ to a representative value of each GMC. Table \ref{table:appendix_clouds} of Appendix \ref{section:appendix} lists the parameters used for the measurement of the upper limits for each GMC without PN detection. Figs. \ref{figure:cloud1} to \ref{figure:cloud10} show the calculated upper limits for the PN column density in these regions as well.

\begin{table}
\tabcolsep1.5pt
\begin{center}
\begin{threeparttable}
\caption{Physical parameters derived for the emission from the molecules analysed in this work towards GMCs 6 and 4 of NGC 253.}
\begin{tabular}{ c c c c c c } 
\hline
%  \noalign{\vskip 3pt}    
\multirow{2}{4em}{\centering Mol.} & \textit{N}$\times$10$^{13}$   & \textit{T$_{\rm ex}$}  & v$_{\rm LSR}$  & FWHM & $\chi$ \\ 
 &  (cm$^{-2}$)  & (K) & (km s$^{-1}$) & (km s$^{-1}$) &  ($\times$10$^{-11}$)\tnote{a}\\
  \hline
  \noalign{\vskip 4pt}
  \multicolumn{6}{c}{GMC 6} \\ [2pt]
  \hline
  \noalign{\vskip 4pt}
  PN & 1.20$\pm$0.17 & 4.4$\pm$1.3 & 184$\pm$6 & 50\tnote{b} & 0.80$\pm$0.19\\
  PO & $<$1.5 & 4.4\tnote{b} & 184.2\tnote{b} & 50\tnote{b} & $<$1.0\\
  PH$_{3}$ & $<$3.2 & 4.4\tnote{b} & 184.2\tnote{b} & 50\tnote{b} & $<$2.2\\
  CP & $<$3.0 & 4.4\tnote{b} & 184.2\tnote{b} & 50\tnote{b} & $<$2.0\\
  CCP & $<$3.8 & 4.4\tnote{b} & 184.2\tnote{b} & 50\tnote{b} & $<$2.6 \\
  %SiO & 32$\pm$4 & 6.29$\pm$0.15 & 185.2$\pm$0.6 & 70.4$\pm$1.4 & 22$\pm$5\\
  $^{29}$SiO & 9.2$\pm$1.8 & 5.2$\pm$0.3 & 187.4$\pm$2.0 & 70.4\tnote{b} & 6.3$\pm$1.9\\
  C$^{34}$S & 57$\pm$7 & 6.83$\pm$0.21 & 188.1$\pm$0.4 & 67.8$\pm$1.0 & 39$\pm$9\\ [2pt]
%  C$_2$H & 1398$\pm$112 & 6.4$\pm$0.2 & 183.7$\pm$0.8 & 55.4$\pm$1.6 & 954$\pm$124\\ [2pt]
  \hline
  \noalign{\vskip 4pt}
  \multicolumn{6}{c}{GMC 4} \\ [2pt]
  \hline
  \noalign{\vskip 4pt}
  PN & 0.65$\pm$0.17 & 4.4\tnote{b} & 242\tnote{b} & 50\tnote{b} & 0.44$\pm$0.17 \\
  PO & $<$1.1 & 4.4\tnote{b} & 242\tnote{b} & 50\tnote{b} & $<$0.7\\
  PH$_{3}$ & $<$3.4 & 4.4\tnote{b} & 242\tnote{b} & 50\tnote{b} & $<$2.3\\
  CP & $<$2.7 & 4.4\tnote{b} & 242\tnote{b} & 50\tnote{b} & $<$1.8\\
  CCP & $<$2.8 & 4.4\tnote{b} & 242\tnote{b} & 50\tnote{b} & $<$1.9\\
  %SiO & 21$\pm$3 & 6.2$\pm$0.3 & 243.4$\pm$0.9 & 53.2$\pm$2.0 & 14$\pm$4 \\ 
  $^{29}$SiO & 6.1$\pm$1.0 & 5.20$\pm$0.23 & 246.4$\pm$1.8 & 74$\pm$4 & 4.1$\pm$1.2 \\
  C$^{34}$S & 25$\pm$3 & 7.9$\pm$0.5 & 237.7$\pm$1.0 & 74.4$\pm$2.3 & 17$\pm$5 \\ [2pt]
  \hline
\end{tabular}
\begin{tablenotes}
   \item[a] \textit{N$_{\rm H_2}$} taken from \cite{Mangum2019}.
   \item[b] Parameter fixed. %Taken from PN in GMC 6.
  \end{tablenotes}
\label{table:parameters}
\end{threeparttable}
\end{center}
\end{table}

\subsection{Non-LTE analysis of PN}
\label{sec:non-LTE}

\begin{figure}
    \centering
    \includegraphics[width=0.5\textwidth]{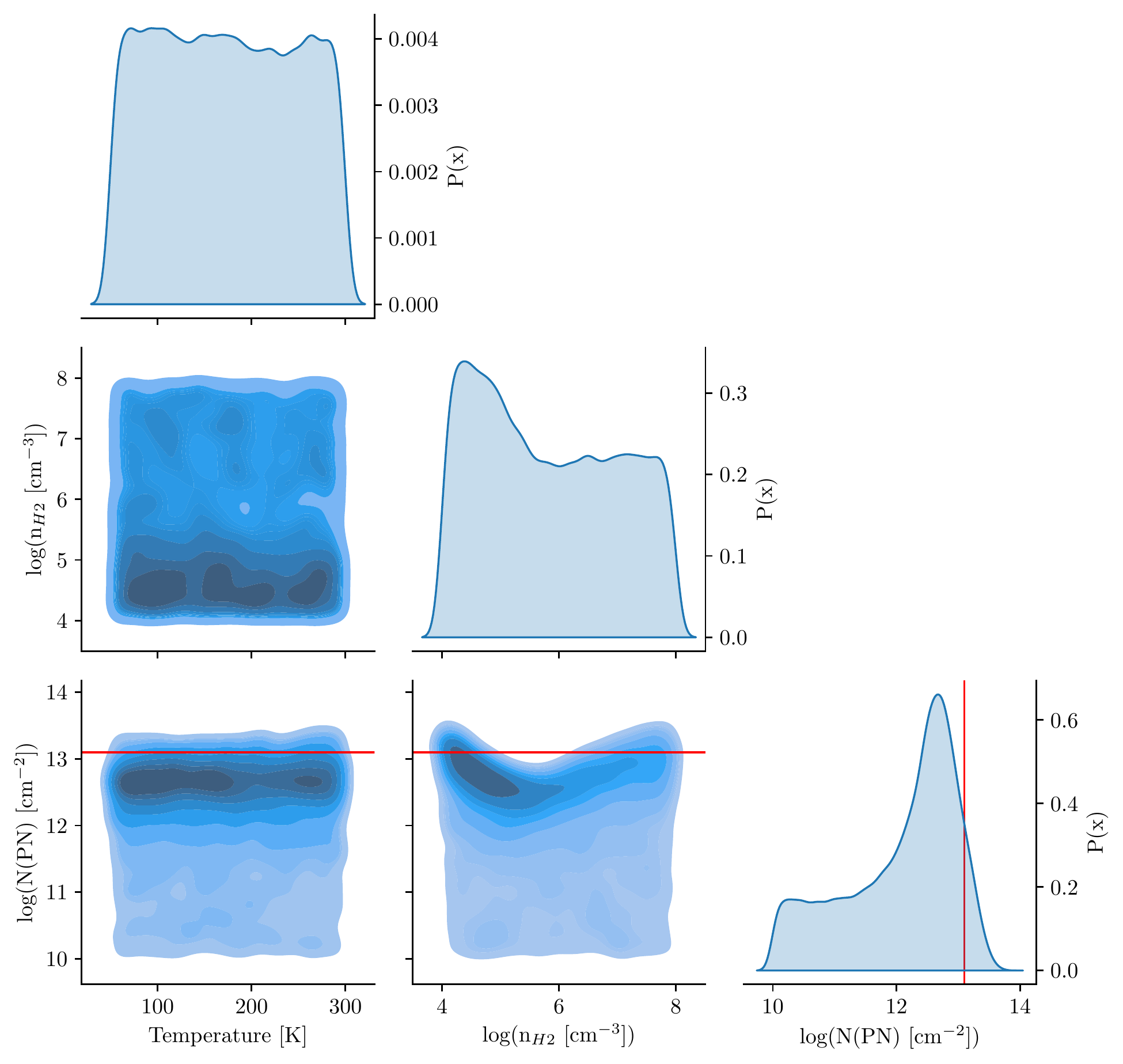}
    \caption{Marginalised and joint posterior distributions from the MCMC sampling. The 1D distributions show the probability a parameter takes a given value, marginalised over all possible combinations of the other parameters. The 2D distributions are darker where a parameter combination is more likely. Column density, volume density and probability scales are logarithmic. The red lines indicate the value of the PN column density obtained with the LTE analysis.}
\label{figure:probs}
\end{figure}

\begin{figure*}
\centering
\includegraphics[width=\textwidth]{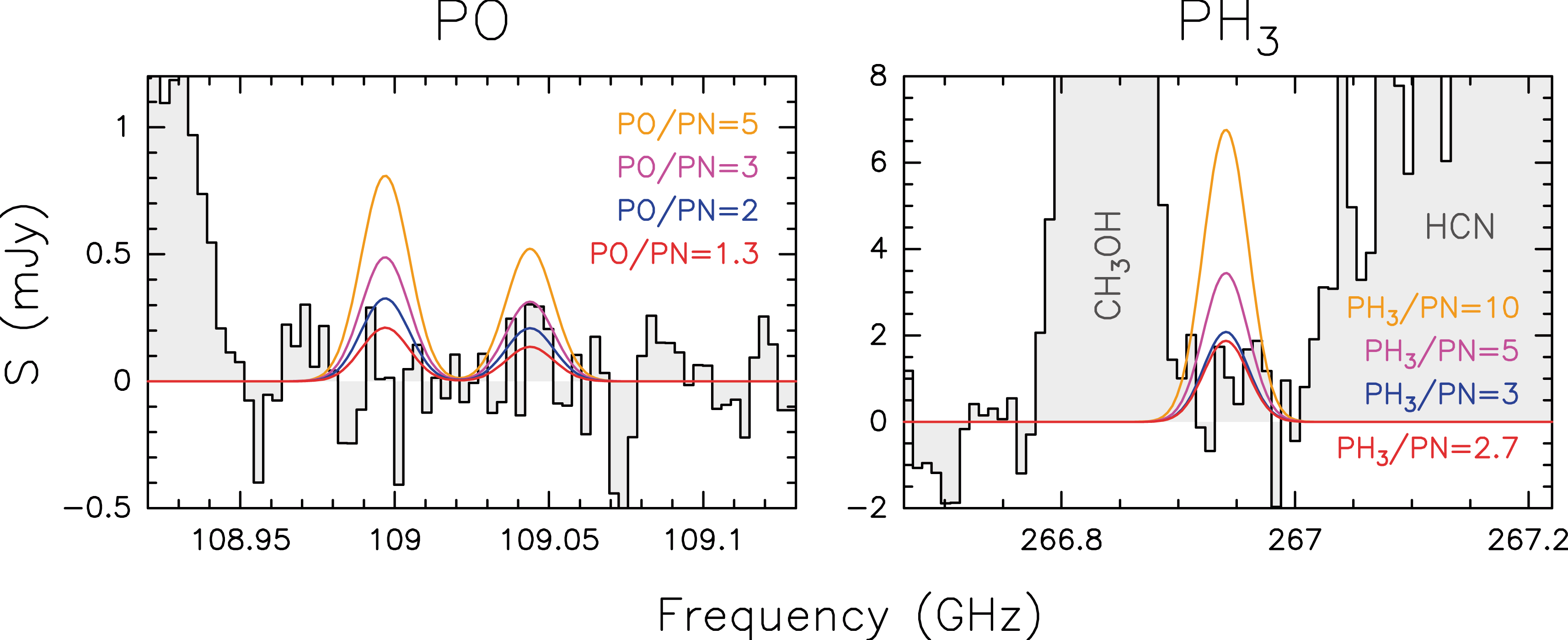}
\caption{{\it Left}: Spectrum towards GMC 6 containing the two PO transitions indicated in Table \ref{table:spectroparam}. The coloured curves show LTE synthetic models assuming different PO/PN ratios, using the $N$ and the $T_{\rm ex}$ obtained for PN. The red curve corresponds to the calculated upper limit for PO. {\it Right}: Same as in left panel, but for PH$_{3}$ (1-0).}
\label{figure:POPH3}
\end{figure*}

We also performed a non-LTE analysis in GMC 6 using SpectralRadex\footnote{\url{https://spectralradex.readthedocs.io}} \citep{Holdship2021}.
SpectralRadex uses RADEX \citep{VanderTak2007} to calculate spectral line excitation and optical depth within the large velocity gradient approximation.
A Gaussian spectral line profile is assumed to produce model spectra.
We used the collisional data between PN and p-H$_2$ \citep{Najar2017PNData} taken from BASECOL \citep{Dubernet2013Basecol,Ba2020basecol} and prepared for RADEX with SPECTCOL\footnote{\url{http://www.vamdc.org/activities/research/software/spectcol/}}.\par
Since SpectralRadex works with intensities in units of K, we converted the intensities using Equation 3.31 in the ALMA Cycle 8 technical Handbook\footnote{\url{https://almascience.nrao.edu/documents-and-tools/cycle8/alma-technical-handbook/view}}:

\begin{equation}
    T (K) = 13.6 \, \left(\frac{300 \, \rm GHz}{\nu} \right)^2 \frac{1}{\theta_{max}('') \, \theta_{min}('')} \, I({\rm Jy \, beam^{-1}})
\end{equation}
where $T$ is the brightness temperature, $I$ is the intensity, and $\theta_{min}$ and $\theta_{max}$ are the axes of the synthesised beam.
This allows us to model the spectra as a function of the PN column density, gas temperature and density. We combine this model with emcee \citep{ForemanMackey2013emcee} to perform a Markov Chain Monte Carlo (MCMC) sampling of the likelihood.  We assume the uncertainty on each channel is independent from the other channels, Gaussian, and 15\% of the channel value. The FWHM and v$_{\rm LSR}$ of the lines are taken from Table~\ref{table:parameters}. \par

The marginalised and joint probability distributions of these parameters are shown in Fig.~\ref{figure:probs}. 
We find that the gas temperature is poorly constrained, showing a range for $T_{\rm K}$ between $\sim$50K to $\sim$300K. Previous measurements \citep{Mangum2013,Mangum2019} obtained kinetic temperatures between 50 K and $>$150 K on size scales as large as 5$''$ in GMC 6.  On smaller scales ($\sim$1$''$) $T_{\rm K}$ could be as high as 300 K.
The most likely value of the column density of PN is 5$\times$10$^{12}$ cm$^{-2}$. To estimate the uncertainty on this value, we take the region around this most likely value which contains 67\% of the probability density (similar to a 1$\sigma$ interval). This interval is 6.3$\times$10$^{11}$ cm$^{-2}$ to 1.8$\times$10$^{13}$ cm$^{-2}$.
The non-LTE analysis provides us with results that are similar to those seen in the LTE analysis. The LTE value of \textit{N}=1.2$\times$10$^{13}$ cm$^{-2}$ (red lines in Fig.~\ref{figure:probs}) falls at the high side of the confidence interval and close to the most likely value of $N$ in the non-LTE analysis (5$\times$10$^{12}$ cm$^{-2}$).

This non-LTE analysis gives a most likely gas density of 3$\times$10$^{4}$ cm$^{-3}$ and the majority of the probability density is at values lower than 10$^{6}$ cm$^{-3}$.

\subsection{PO (phosphorus monoxide)}

We searched for PO using spectroscopic data from the CDMS catalogue entry 047507 (October 2019) based on the works by \cite{Bailleux2002} and the dipole moment from \cite{Kanata1988} (see Table \ref{table:spectroparam}). PO is not detected in any of the GMCs. 
We show in the left panel of Fig. \ref{figure:POPH3} the spectrum towards GMC 6 (that with the brightest PN detection). This spectral window contains the two PO transitions indicated in Table \ref{table:spectroparam}, corresponding to the $J$=5/2$-$3/2 $\Omega$=1/2 quadruplet (the other two lines of this quadruplet are heavily contaminated by other molecular species). We derived an upper limit for PO using the transitions shown in Fig. \ref{figure:POPH3}, assuming the FWHM and $T_{\rm ex}$ derived from the LTE analysis of PN. 
We obtain \textit{N}$<$1.5$\times$10$^{13}$ cm$^{-2}$ for GMC 6 (Table \ref{table:parameters}), which gives a PO/PN ratio $<$1.3.
The same was made for GMC 4 which results in \textit{N}$<$1.1$\times$10$^{13}$ cm$^{-2}$ and a PO/PN ratio $<$1.7.

\subsection{PH$_{3}$ (phosphine)}

We searched for PH$_{3}$ using data from CDMS catalogue entry 034501 (May 2013) based on the works by \cite{Muller2013}. 
The only transition of this species targeted by the ALCHEMI survey is the 1$-$0 transition at 266.945 GHz (Table \ref{table:spectroparam}). We show in the right panel of Fig. \ref{figure:POPH3} the spectrum at this frequency towards GMC 6. The transition is not detected and we derived an upper limit for the column density. We obtained \textit{N}$<$3.2$\times$10$^{13}$ cm$^{-2}$ (Table \ref{table:parameters}) for GMC 6, which gives a PH$_3$/PN ratio $<$2.7.
For GMC 4 the results obtained were \textit{N}$<$3.4$\times$10$^{13}$ cm$^{-2}$ and a PH$_3$/PN ratio $<$5.3.

\subsection{Other P-bearing molecules}

We also searched for other P-bearing molecules (CP, CCP, HCP, and PS) towards GMCs 4 and 6, where PN is detected.
The spectroscopic data of CP, CCP, and HCP was taken from CDMS catalogue entry 043501 (March 2000), entry 055503 (April 2009), and entry 044502 (June 2007), respectively.
The spectroscopic data of PS was taken from the entry 63007 (January 1997) of the Jet Propulsion Laboratory (JPL) catalogue\footnote{https://spec.jpl.nasa.gov
}. 
None of these P-bearing species were detected. All of the transitions of HCP and PS are heavily contaminated by other species, which prevents the derivation of significant upper limits for their abundance.
For CP and CCP we derived upper limits for their abundances using uncontaminated transitions listed in Table \ref{table:spectroparam}. 
The parameters \textit{T$_{\rm ex}$}, v$_{\rm LSR}$, and FWHM were fixed to the values of the PN detection in each source. The derived upper limits for CP and CCP are \textit{N}$<$3.0$\times$10$^{13}$ cm$^{-2}$ and \textit{N}$<$3.8$\times$10$^{13}$ cm$^{-2}$, respectively for GMC 6, and \textit{N}$<$2.7$\times$10$^{13}$ cm$^{-2}$ and \textit{N}$<$2.8$\times$10$^{13}$ cm$^{-2}$, respectively for GMC 4 (Table \ref{table:parameters}).

\subsection{Detections of $^{29}$SiO and C$^{34}$S}
\label{sec:other-molecules}

We also analysed the emission of C$^{34}$S and $^{29}$SiO, an optically-thin isotopologue of SiO, which will be used in the discussion below (Sect. \ref{sec:discussion}). 
The LTE fits were done with MADCUBA-AUTOFIT, using the same procedure described in Sect. \ref{sec:analysis-LTE}.
The molecular transitions used in the analysis are listed in Table \ref{table:spectroparam}. The results of the fits are shown in Figs. \ref{figure:cloud4} to \ref{figure:cloud10}, and Table \ref{table:appendix_clouds} lists the derived parameters of the detections.

\begin{figure}
\centering
\includegraphics[width=0.5\textwidth]{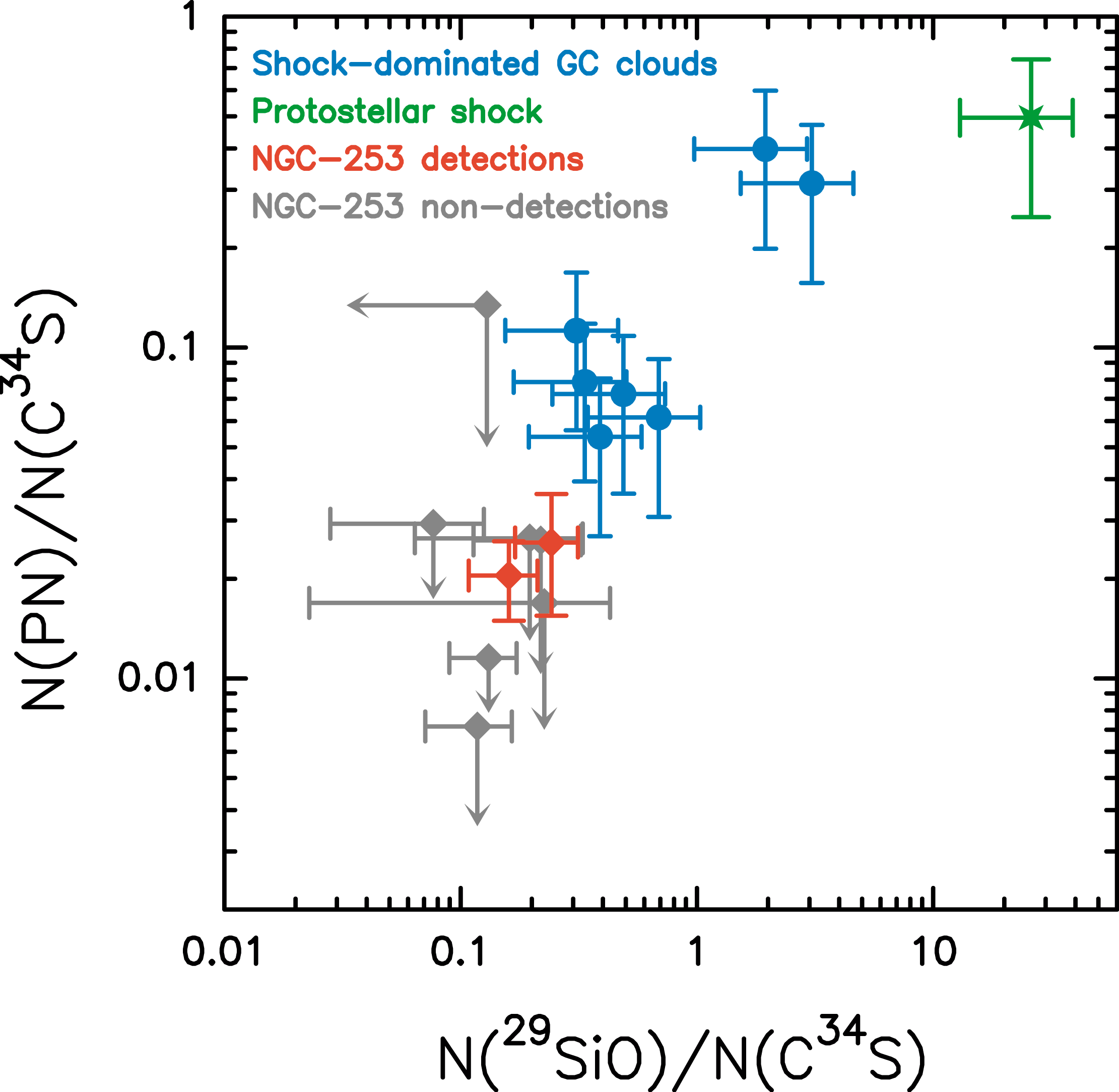}
  \caption{PN and $^{29}$SiO column density ratios relative to C$^{34}$S. The two detections of PN in NGC 253 (GMCs 4 and 6) are indicated with red diamonds, while the non-detections towards the remaining GMCs are indicated with grey diamonds. We also added data from other sources in our Galaxy Centre (\citealt{Rivilla2018} and references therein): in green we plot the L1157-B1 shock data and the blue dots represent data from shock dominated Galactic Centre clouds.}
\label{figure:PNvs29SiO}
\end{figure}

\section{Discussion: Phosphorus chemistry}
\label{sec:discussion}

\begin{table}
\tabcolsep1pt
\begin{center}
\begin{threeparttable}
\caption{PO/PN ratios in  NGC 253 and several regions in our Galaxy.}
\begin{tabular}{ c c c c } 
\hline
%  \noalign{\vskip 3pt}    
Source & [PO/PN] & [PH$_3$/PN] & Region Type \\ %[2pt]
  \hline
%    \noalign{\vskip 4pt}   
\multicolumn{4}{c}{Extragalactic}   \\
\hline
  NGC 253 - GMC 4 & $<$1.7 & $<$5.2 & EGMC \\
  NGC 253 - GMC 6 & $<$1.3 & $<$2.7 & EGMC \\
  \hline
\multicolumn{4}{c}{Galactic Centre}   \\
\hline
  G+0.693-0.027\tnote{1} & 1.5$\pm$0.4 & -- & GCMC \\
  S+0.24+0.01\tnote{1} & $<$2.7 & --& GCMC \\
  M-0.02-0.07\tnote{1} & $<$3.1 &-- & GCMC \\
    \multirow{2}{4em}{\centering SgrB2 N\tnote{1}} & $<$1.8 & -- & \multirow{2}{7em}{\centering MSFR} \\
  & $<$3.2 & -- \\
  \multirow{2}{4em}{\centering SgrB2 M\tnote{1}} & $<$6.8 &--  & \multirow{2}{7em}{\centering MSFR} \\
  & $<$3.5 & --  & \\
  \hline
 \multicolumn{4}{c}{Galactic disk star-forming regions}   \\ 
 \hline
  AFGL5142 P1\tnote{2}      & $<$1.1 & -- & MSFR \\
  AFGL5142 P2\tnote{2} & 0.6$\pm$0.2 / 1.8  & -- & MSFR \\
  AFGL5142 P3\tnote{2} & 1.4$\pm$0.3 / 3.3 &-- & MSFR \\
  AFGL5142 P4\tnote{2} & 1.6$\pm$0.2 / 4.4 & -- & MSFR \\
  AFGL5142 P5\tnote{2} & 2.6$\pm$0.6 / 6.4 & -- & MSFR \\
  AFGL5142 P6\tnote{2} & 2.2$\pm$0.6 / 4.6 & -- & MSFR \\
%  AFGL5142 P7\tnote{2} & $<$0.9 & -- & MSFR \\
%  AFGL5142 MM\tnote{2} & $<$0.8 & -- & MSFR \\
  W51\tnote{3} & 1.9 &--  & MSFR \\
  W3(OH)\tnote{3} & 3 &--  & MSFR \\
  Orion plateau\tnote{4} & 2.7 & -- & MSFR \\
  \multirow{2}{7em}{\centering L1157-B1\tnote{5}} & \multirow{2}{4em}{\centering 2.6} & \multirow{2}{4em}{\centering $<$1.1} & LMSFR \\
   &  &  & (protostellar shock) \\ 
  B1-a\tnote{6} & 1-3 &--  & LMSFR \\ 
%  67P/Churyumov-Gerasimenko\tnote{1} & $>$10 & comet \\ 
  [2pt]
  \hline
  \noalign{\vskip 5pt}
\end{tabular}
\begin{tablenotes}%[para]
    \item Region types are as follows: Extragalactic giant molecular clouds (EGMC), Galactic Centre molecular clouds (GCMC), massive star-forming regions (MSFR), and low-mass star-forming regions (LMSFR).
   \item[1] \cite{Rivilla2018}.
   \item[2] Different spots with PN and PO emission identified towards the massive star-forming region AFGL 5142 by \cite{Rivilla2020}. Two different values are presented: those derived from an LTE analysis (left), and those derived from a non-LTE analysis assuming $T_{\rm kin}$=20 K and gas density of 10$^{5}$ cm$^{-3}$.
   \setlength{\columnsep}{0.8cm}
   \setlength{\multicolsep}{0cm}
   \begin{multicols}{2}
     \item[3] \cite{Rivilla2016}
     \item[4] \cite{Bernal2021}
     \item[5] \cite{Lefloch2016}
     \item[6] \cite{Bergner2019}
  \end{multicols}
  \end{tablenotes}
\label{table:POPNratio}
\end{threeparttable}
\end{center}
\end{table}

\subsection{Correlation between PN and SiO abundances: Role of shocks}
\label{sec:discussion-shocks}

Although the chemistry of P in the ISM is still poorly constrained, the observations of P-bearing species in the last years have suggested the likely importance of shocks. In Galactic sources (Galactic Centre molecular clouds, massive star-forming regions, and protostellar shocks in the Galactic disk), a good correlation has been established between the abundances of SiO $-$ a well known shock tracer $-$ and those of PN (\citealt{Rivilla2018}). 
Following this work, we show in Fig. \ref{figure:PNvs29SiO} the column densities of PN and $^{29}$SiO, both normalised by the column density of C$^{34}$S, towards different Galactic sources detected previously. \citet{Rivilla2018} used C$^{34}$S to normalise the abundances because this optically-thin species is expected to be a good proxy of H$_2$, and it is barely affected by the presence of shocks or photo-dominated regions \citep{Requena-Torres2006,martin2009}.
We include in Fig. \ref{figure:PNvs29SiO} the two points corresponding to the PN detections in GMCs 4 and 6, and the upper limits derived for the rest of the GMCs. 
The points of NGC 253 nicely follow the same trend observed in Galactic sources, extending it to lower relative abundances.

This observational evidence suggests that the chemistry of P in extragalactic environments is overall similar to that in our Galaxy. The emission of P-bearing molecules is enhanced in shocked regions, where shock-induced sputtering of grains injects the P locked on the grain surfaces into the gas phase \citep{Rivilla2020}. The main carrier of P on the icy mantles has been suggested to be PH$_3$, which can be rapidly formed by hydrogenation of atomic P, and that can be converted to PN and PO in the gas phase (see \citealt{Jimenez-Serra2018}).
We note that the only interferometric maps of P-bearing emission available so far, obtained towards the star-forming region AFGL 5142 (\citealt{Rivilla2020}), have shown that the emission of P-bearing molecules arises from regions associated with shocks, and not from the central hot molecular core surrounding the newly formed protostars, which clearly rule out desorption due to heating.

\subsection{Molecular ratios of P-bearing species: Role of cosmic-rays and photochemistry}
\label{sec:discussion-ratios}

It is still not fully understood what the main carrier of P in the ISM is.
The pioneering chemical models by \cite{Thorne1987} proposed PO as the most abundant P-bearing molecule in the ISM. However, subsequent studies by \cite{Millar1987} and \cite{Adams1990} indicated that PO is expected to have a lower abundance than that of PN in the ISM. 
More recently, the chemical models by \citet{Jimenez-Serra2018} proposed that PH$_3$ could be the main carrier of P on the surface of dust grains due to rapid hydrogenation of atomic P, and hence that it might be also relatively abundant in the gas phase if efficiently desorbed.
In the last years, several observations of P-bearing species have allowed us to quantify the relative abundances of the different species, which are summarised in Table \ref{table:POPNratio} (see references therein). 
In Galactic star-forming environments (low- and high-mass star-forming regions and protostellar shocks), the PO/PN ratio found is usually $>$1, with values up to $\sim$6 (Table \ref{table:POPNratio}). 
In molecular clouds of the CMZ of our Galaxy, the only detection of PO provides PO/PN ratio of 1.5$\pm$0.4 (Table \ref{table:POPNratio}).
In the GMCs 4 and 6 of NGC 253 we found that the PO/PN ratio is $<$1.7 and $<$1.3, respectively.

\citet{Rivilla2020} found in the AFGL 5142 star-forming region that the PO/PN ratios in different shocked regions decrease with decreasing distance to the source of ultraviolet (UV) photons (the central protostar). These authors suggested that the abundance of P-bearing molecules could be regulated by the photo-destruction if PO is more efficiently photo-destroyed than PN, as proposed by \citet{Jimenez-Serra2018}. This might explain why the searches of PO towards the CMZ of NGC 253 and our Galaxy resulted in non-detections, with the only exception of the molecular cloud G+0.693-0.027 (\citealt{Rivilla2018}). The CMZ of galaxies harbor high cosmic-ray ionisation rates ($\zeta$). \citet{goto2014} found that $\zeta$ is well above 10$^{-15}$ s$^{-1}$ in the centre of our Galaxy, several orders of magnitude higher than the standard Galactic value of $\zeta\sim$1.3$\times$10$^{-17}$ s$^{-1}$ (e.g. \citealt{Padovani2009}). In the CMZ of NGC~253, although the values of $\zeta$ are not well constrained yet, recent ALCHEMI-based research has shown that they are between 10$^{-11}$ and 10$^{-14}$ s$^{-1}$ (\citealt{Holdship2021,harada2021}). 
This high $\zeta$ produces a field of secondary ultraviolet (UV) photons, which might result in the aforementioned more efficient photo-destruction of PO than PN. %\citep{Jimenez-Serra2018,Rivilla2020}.
If this is the case, the abundance of PO would significantly decrease in galactic CMZs, producing values of PO/PN$<$1.

In any case, we note that the upper limits derived for the PO abundance towards the CMZ of NGC 253 and other regions in the CMZ of our Galaxy (Table \ref{table:POPNratio}) do not allow us to firmly confirm that the ratio PO/PN$<$1. Deeper observations, both in the Galactic disk and towards more molecular clouds in the CMZ of our Galaxy and other galaxies, are needed to confirm this hypothesis.

Regarding PH$_3$, we found that PH$_3$/PN is $<$5.2 and $<$2.7 in GMCs 4 and 6, respectively. These upper limits indicate that PH$_3$ is not significantly more abundant than PN in the gas phase, confirming the previous result by \citet{Lefloch2016}, who found that PH$_3$/PN$<$1.1 in a Galactic protostellar shock (Table \ref{table:POPNratio}). As for the case of PO, deeper observations will be needed to confirm whether PH$_3$ are equally or less abundant than PN in NGC 253.

\subsection{Comparison with chemical models}
\label{sec:discussion-chemistry}

Based on the observational evidence, we have discussed in the Sections \ref{sec:discussion-shocks} and \ref{sec:discussion-ratios} the possible role of shocks and cosmic rays (+ photochemistry) shaping the molecular abundances of P-bearing species in the ISM. 
In this section, we compare the observational results
with the predictions of the chemical models presented by \cite{Jimenez-Serra2018}, who studied the chemistry of P-bearing molecules under different physical conditions such as presence of shocks, and different cosmic-ray ionisation rates ($\zeta$) and gas densities.

The shock models predict high gas-phase abundances of P-bearing species, which are significantly enhanced due to grain sputtering.
For PN, its abundance is (1$-$5)$\times$10$^{-10}$, which is more than one order of magnitude higher than the abundances we derived towards the GMCs of NGC 253 of (0.4$-$0.8)$\times$10$^{-11}$ (assuming that PN and dust trace similar gas). 
Therefore, the models based solely on shocks fail to explain the relatively low PN abundances observed in NGC~253, and hence an additional agent is needed. 

As discussed in Sect. \ref{sec:discussion-ratios}, cosmic rays can also strongly affect the chemistry of P-bearing molecules.
The secondary UV-photon radiation field induced by an enhancement of $\zeta$ can also partially photodissociate the P-bearing molecules. This will regulate the PO/PN ratio (Sect. \ref{sec:discussion-ratios}), but will also decrease the overall molecular abundances. 
When \cite{Jimenez-Serra2018} introduced a high cosmic-ray ionisation rates of $\zeta$=10$^{-13}$ s$^{-1}$ in the shock models with pre-shock densities of $n$(H)=2$\times$10$^{4}$ cm$^{-3}$ (see their Fig. 8), the predicted PN abundances decrease by one order of magnitude at the post-shock stage, reaching several 10$^{-11}$, which is more similar to the ones we derived for NGC 253.
Moreover, we note that these models predict a PO/PN ratio $<$ 1, which is consistent with the upper limits found in the CMZs of NGC 253.
These values for $\zeta$ and gas density match well with the expected physical properties of the GMCs in the CMZ of NGC 253.
As mentioned in Sect. \ref{sec:discussion-ratios}, recent ALCHEMI works indicate that $\zeta$ is 10$^{-11}-$10$^{-14}$ s$^{-1}$ (\citealt{Holdship2021,harada2021}), while the non-LTE analysis of PN (Sect. \ref{sec:non-LTE}) shows that the most likely gas density is around $n$(H$_2$)=3$\times$10$^4$ cm$^{-3}$, and thus\footnote{We note that the non-LTE analysis uses the gas density of molecular hydrogen, $n$(H$_2$), while the chemical models use the gas density of atomic hydrogen, $n$(H). Assuming a molecular cloud in which all the hydrogen is in molecular form, $n$(H)=2$\times n$(H$_2$).} $n$(H)=6$\times$10$^4$ cm$^{-3}$.

\section{Conclusions}

We searched for P-bearing molecules in the Central Molecular Zone of the starburst Galaxy NGC 253 using the unbiased ALCHEMI spectral survey. Phosphorus nitride (PN) was detected towards two of the Giant Molecular clouds (GMCs 4 and 6) in NGC 253, being the first detections of a P-bearing molecule in an extragalactic environment. An LTE analysis yields total PN beam-averaged column densities of $N$=(1.20$\pm$0.09)$\times$10$^{13}$ cm$^{-2}$ and $N$=(6.5$\pm$1.6)$\times$10$^{12}$ cm$^{-2}$ for GMCs 6 and 4, respectively, and an excitation temperature of $T_{\rm ex}$=(4.4$\pm$1.3) K towards GMC 6. The derived molecular abundances with respect to H$_2$ are $\chi$=(8.0$\pm$1.0)$\times$10$^{-12}$ and $\chi$=(4.4$\pm$1.2)$\times$10$^{-12}$ for GMCs 6 and 4, respectively.
A non-LTE analysis derives a column density consistent with the LTE values, and confirms that the PN transitions are sub-thermally excited in a gas with volume density of the order of 10$^4$ cm$^{-3}$. 
Other P-bearing molecules like PO, PH$_{3}$, CP, and CCP were not detected. We provide upper limits for their column densities in GMCs 4 and 6. 
The derived PO/PN ratios are $<$1.3 and $<$1.7 for GMCs 4 and 6, respectively.
  
The comparison between the current observational evidence of P-bearing molecules in Galactic and extragalactic environments and the available chemical models 
indicates that the chemistry of P-bearing species is mainly regulated by shocks, which contribute to desorb the P-content locked on the dust via grain sputtering, and subsequent gas-phase photochemistry that can regulate the molecular abundances of P-bearing species and the PO/PN ratio. This photochemistry can be triggered by the presence of an intense cosmic-ray flux, as expected in the CMZ of galaxies, or by a nearby protostar in star-forming regions.

\begin{acknowledgements}

We acknowledge the anonymous reviewer for their suggestions that have improved the original manuscript.
The authors thank Marie-Lise Dubernet and Yaye Awa Ba for providing us with the PN collisional data ahead of the latest BASECOL release.
This paper makes use of the following ALMA data: ADS/JAO.ALMA\#2017.1.00161.L and ADS/JAO.ALMA\#2018.1.00162.S. ALMA is a partnership of ESO (representing its member states), NSF (USA) and NINS (Japan), together with NRC (Canada), MOST and ASIAA (Taiwan), and KASI (Republic of Korea), in cooperation with the Republic of Chile. The Joint ALMA Observatory is operated by ESO, AUI/NRAO and NAOJ.  The National Radio Astronomy Observatory is a facility of the National Science Foundation operated under cooperative agreement by Associated Universities Inc.
V.M.R. and L.C. have received funding from the Comunidad de Madrid through the Atracci\'on de Talento Investigador (Doctores con experiencia) Grant (COOL: Cosmic Origins Of Life; 2019-T1/TIC-15379). J.H. and S.V. are funded by the European Research Council (ERC) Advanced Grant MOPPEX 833460.
K.N. is supported JSPS KAKENHI Grant Number 19K03937.
\end{acknowledgements}


\bibliographystyle{aa}
\bibliography{aanda}

\begin{appendix}
\section{Additional molecular line analysis towards NGC 253 GMCs}
\label{section:appendix}

\begin{figure}[h]
\centering
\includegraphics[width=8.8cm]{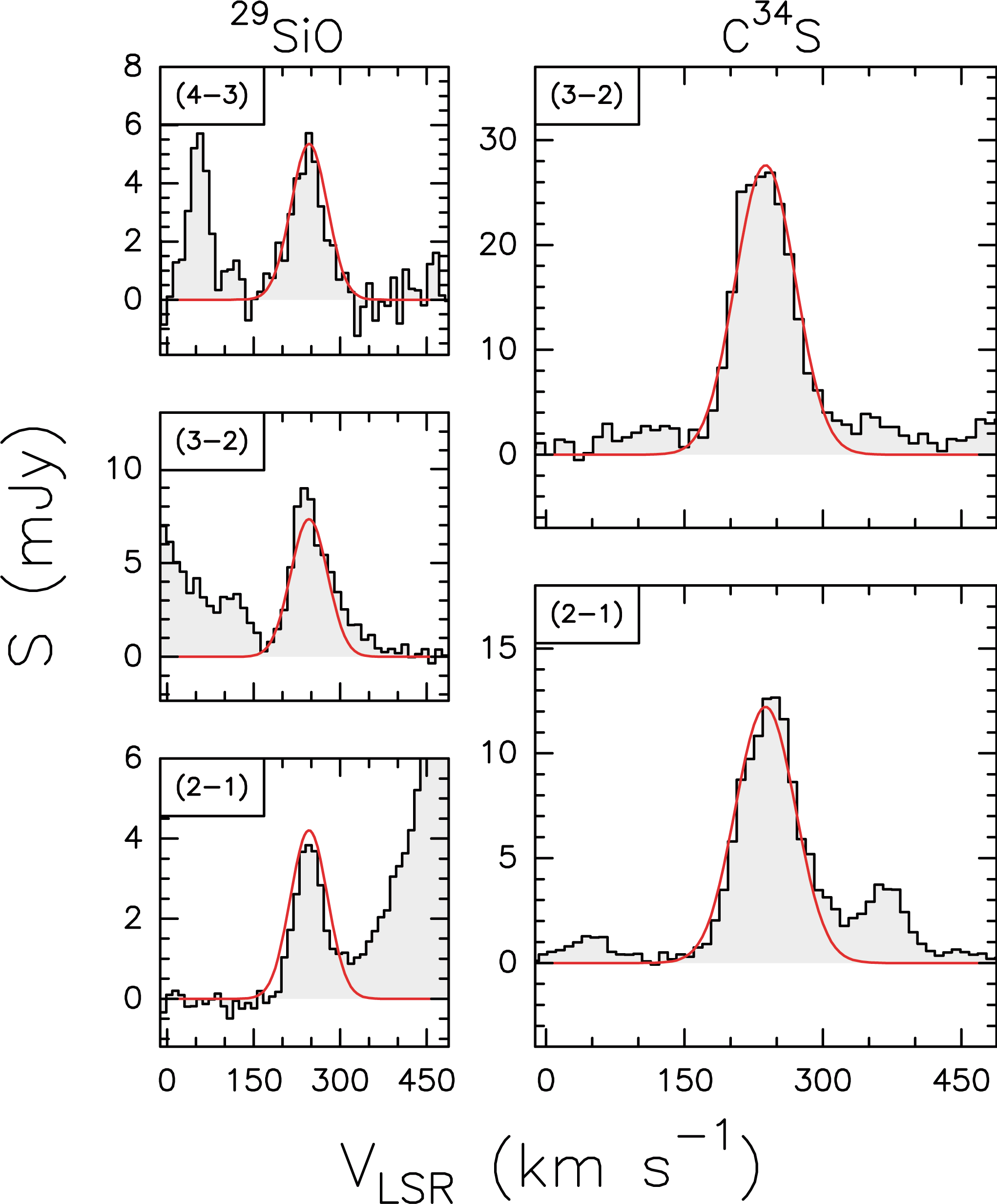}
  \caption{Spectra of $^{29}$SiO and C$^{34}$S towards GMC 4 are shown with grey histograms. The transition is indicated in the upper left of each panel. The red lines indicate the LTE best fits.}
     \label{figure:cloud4}
\end{figure}

\begin{figure}
\centering
\includegraphics[width=8.8cm]{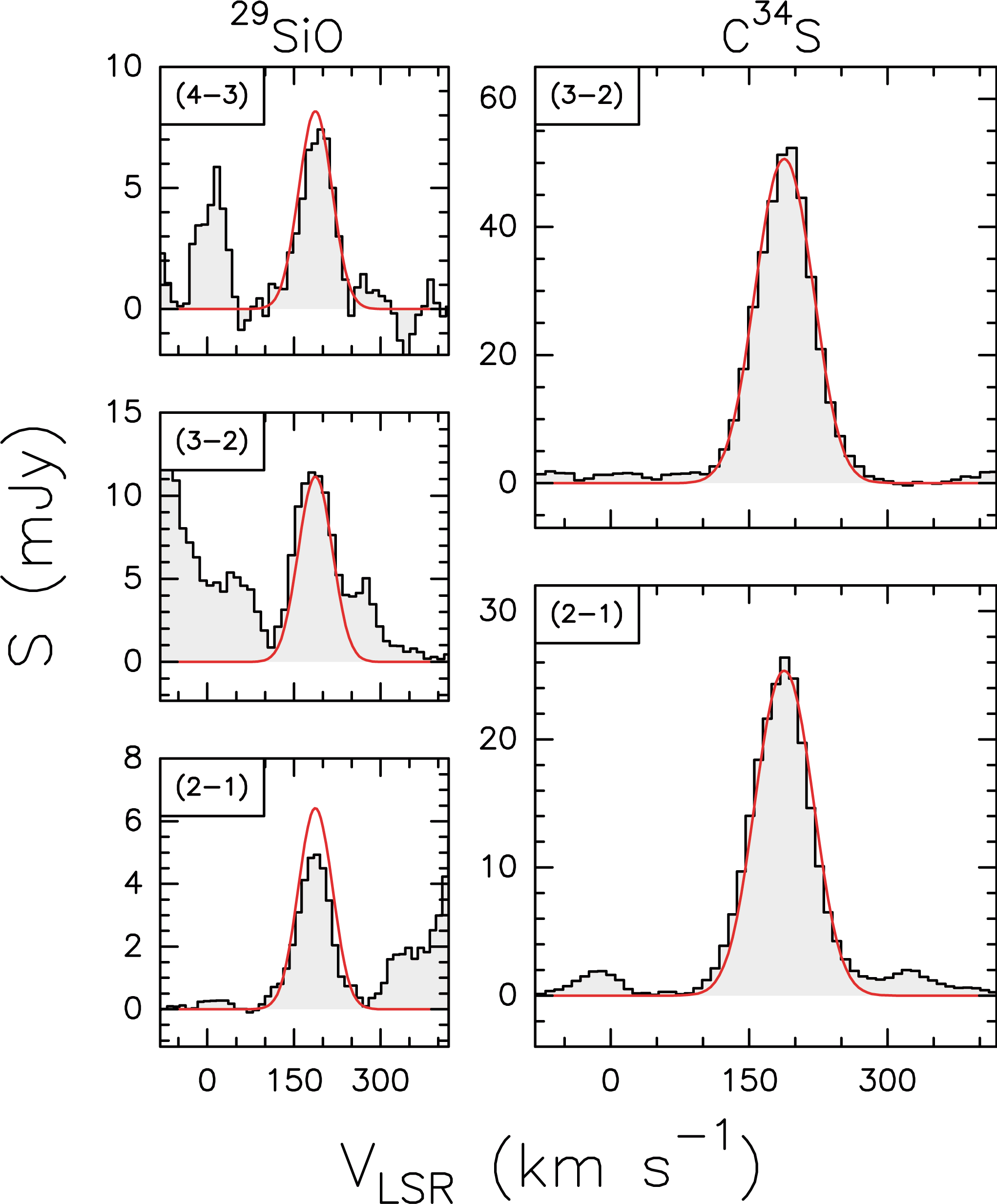}
  \caption{Spectra of $^{29}$SiO and C$^{34}$S towards GMC 6 are shown with grey histograms. The transition is indicated in the upper left of each panel. The red lines indicate the LTE best fits.}
     \label{figure:cloud6}
\end{figure}

\begin{figure*}
\centering
\includegraphics[width=14cm]{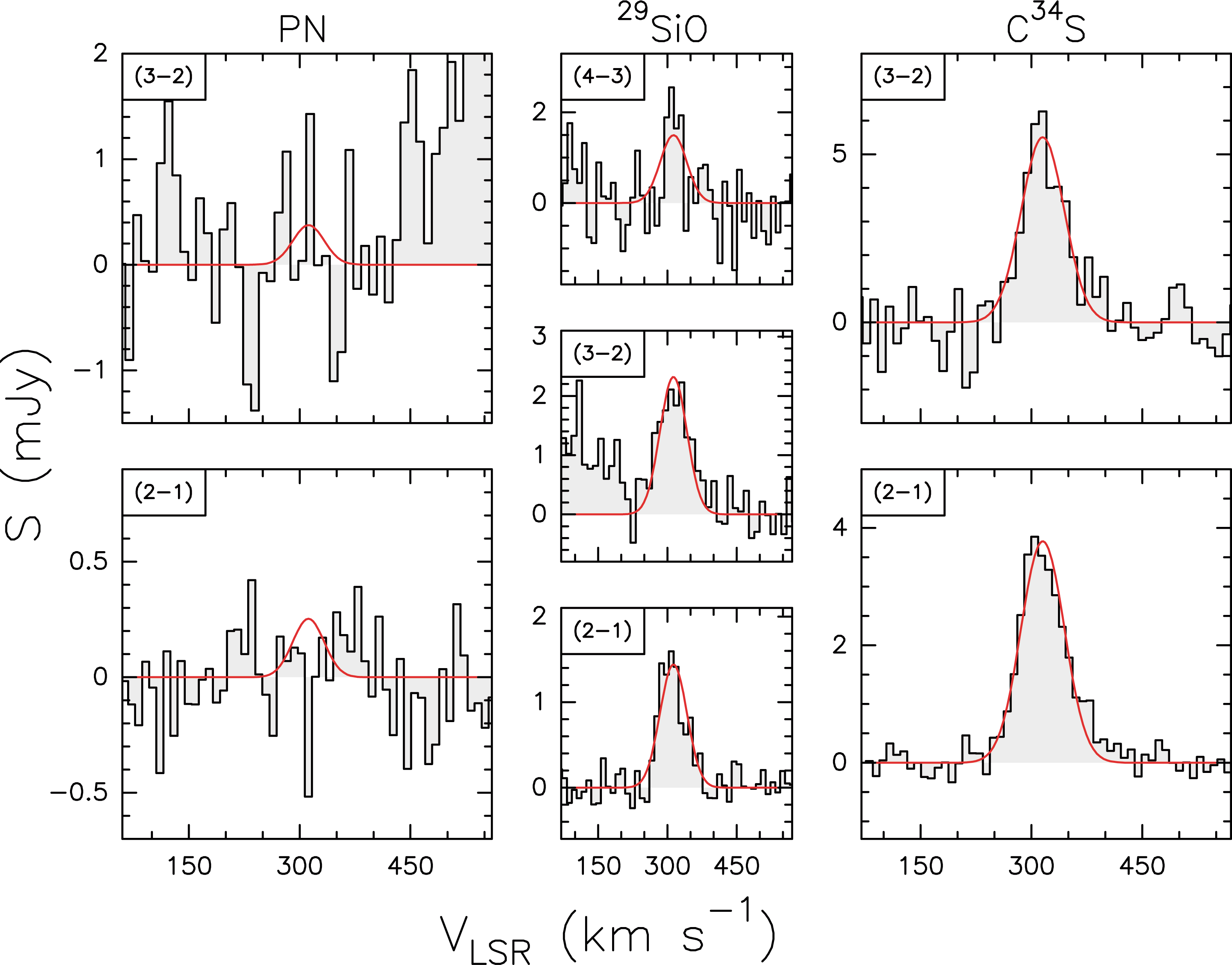}
  \caption{Spectra of PN, $^{29}$SiO, and C$^{34}$S towards GMC 1 are shown with grey histograms. The transition is indicated in the upper left of each panel. The red lines indicate the LTE best fits except for PN, for which the red line indicates an upper limit.}
     \label{figure:cloud1}
\end{figure*}

\begin{figure*}
\centering
\includegraphics[width=14cm]{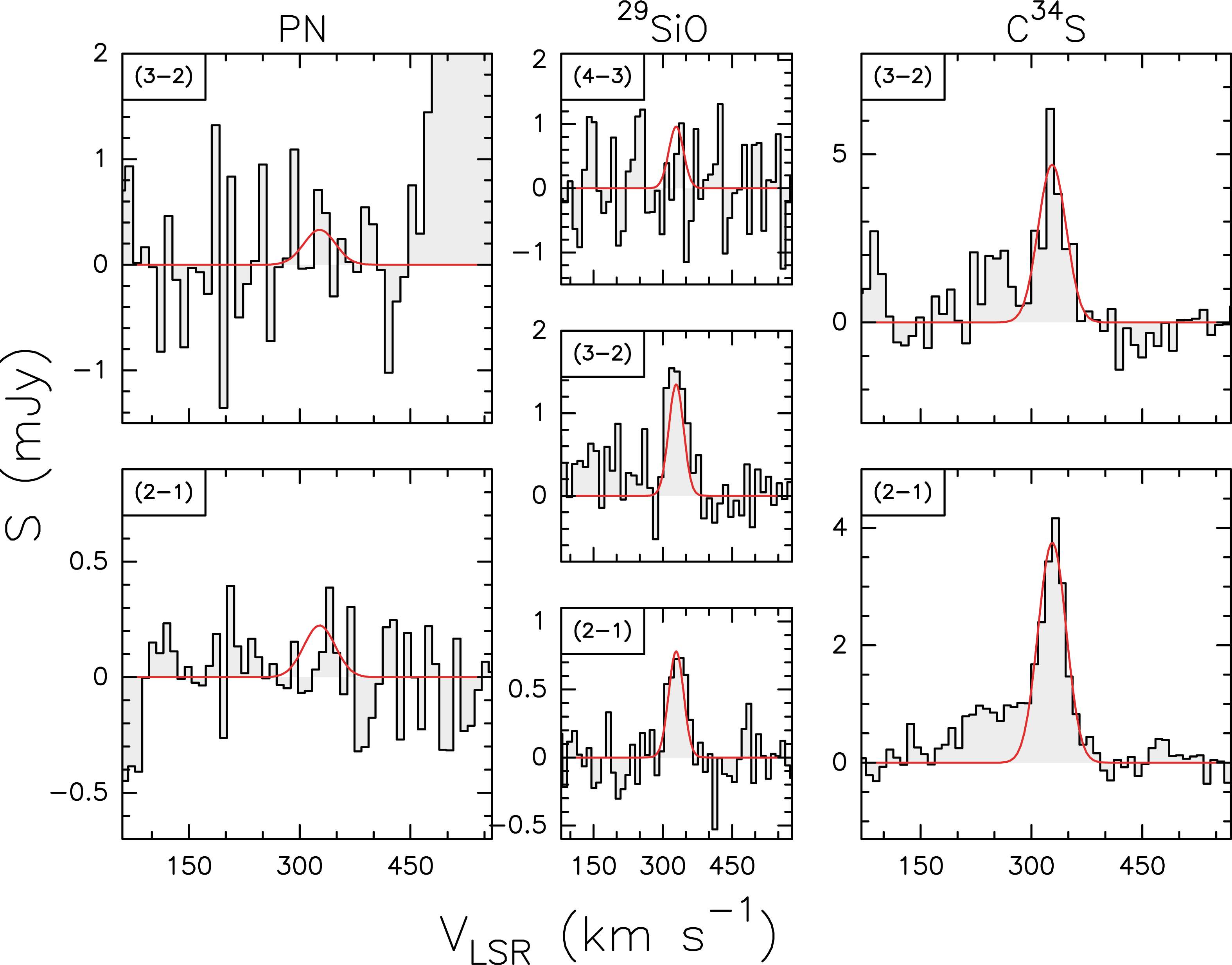}
  \caption{Spectra of PN, $^{29}$SiO, and C$^{34}$S towards GMC 2 are shown with grey histograms. The transition is indicated in the upper left of each panel. The red lines indicate the LTE best fits except for PN, for which the red line indicates an upper limit.}
     \label{figure:cloud2}
\end{figure*}

\begin{figure*}
\centering
\includegraphics[width=14cm]{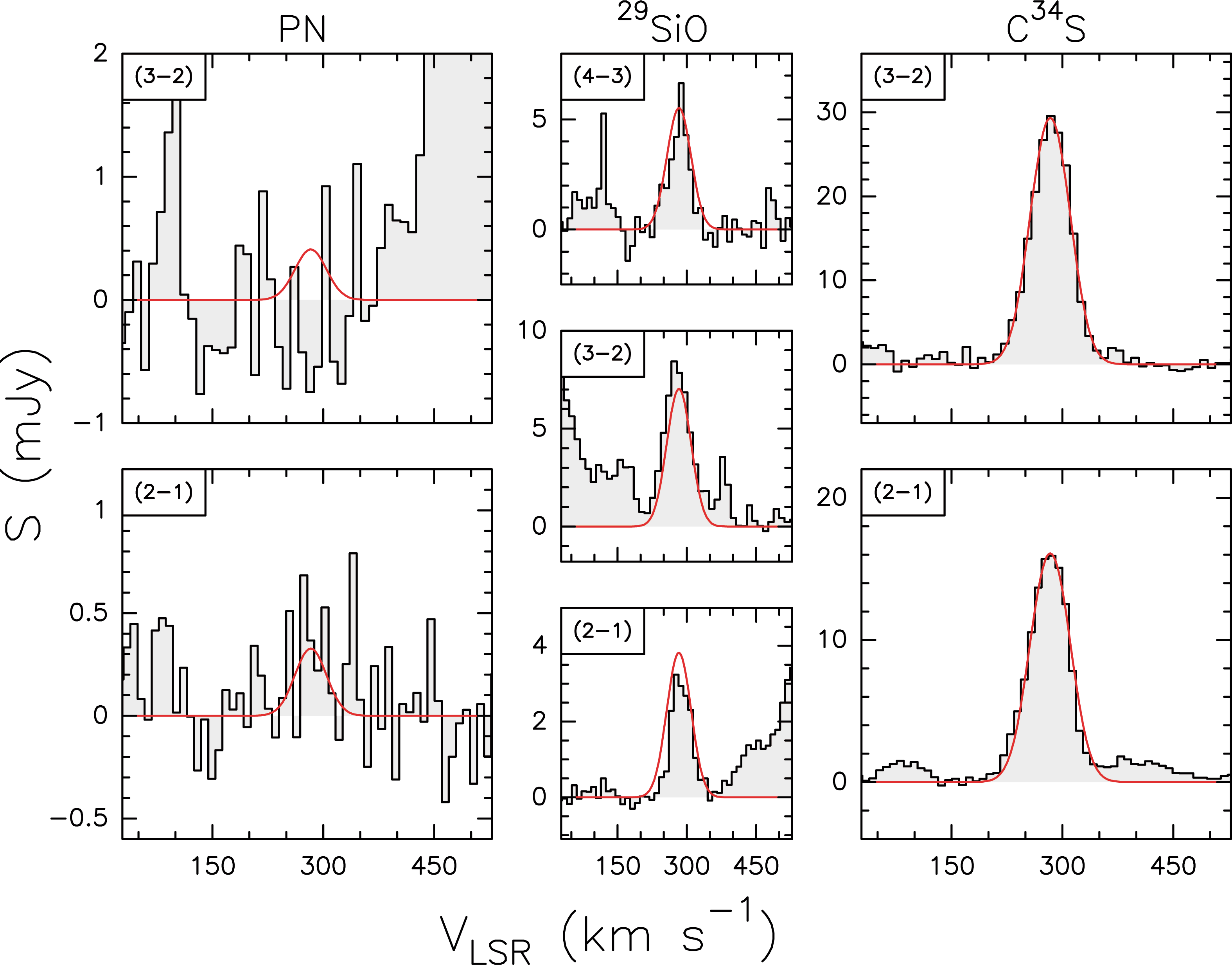}
  \caption{Spectra of PN, $^{29}$SiO, and C$^{34}$S towards GMC 3 are shown with grey histograms. The transition is indicated in the upper left of each panel. The red lines indicate the LTE best fits except for PN, for which the red line indicates an upper limit.}
     \label{figure:cloud3}
\end{figure*}

\begin{figure*}
\centering
\includegraphics[width=14cm]{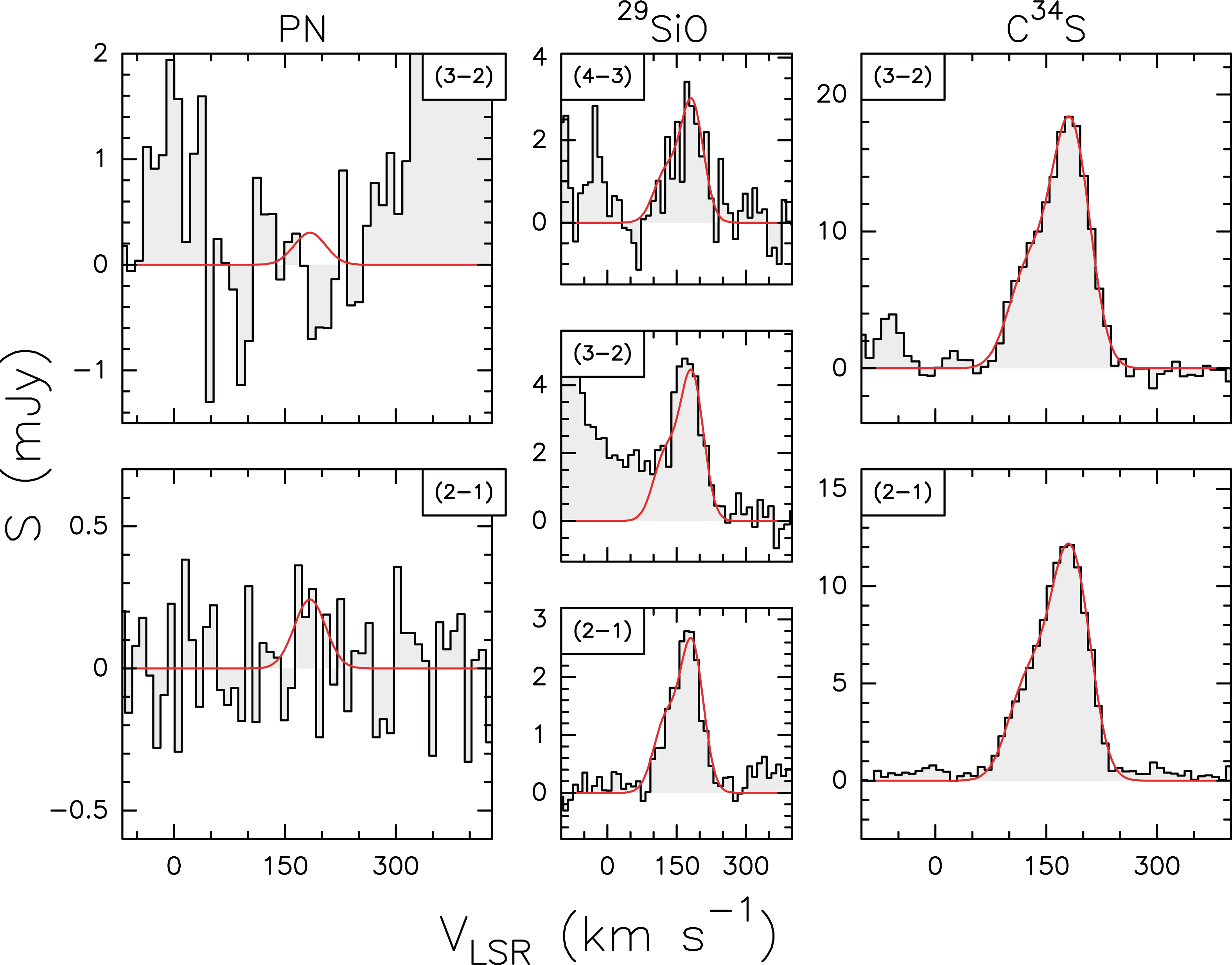}
  \caption{Spectra of PN, $^{29}$SiO, and C$^{34}$S towards GMC 7 are shown with grey histograms. The transition is indicated in the upper left of each panel. The red lines indicate the LTE best fits except for PN, for which the red line indicates an upper limit.}
     \label{figure:cloud7}
\end{figure*}

\begin{figure*}
\centering
\includegraphics[width=14cm]{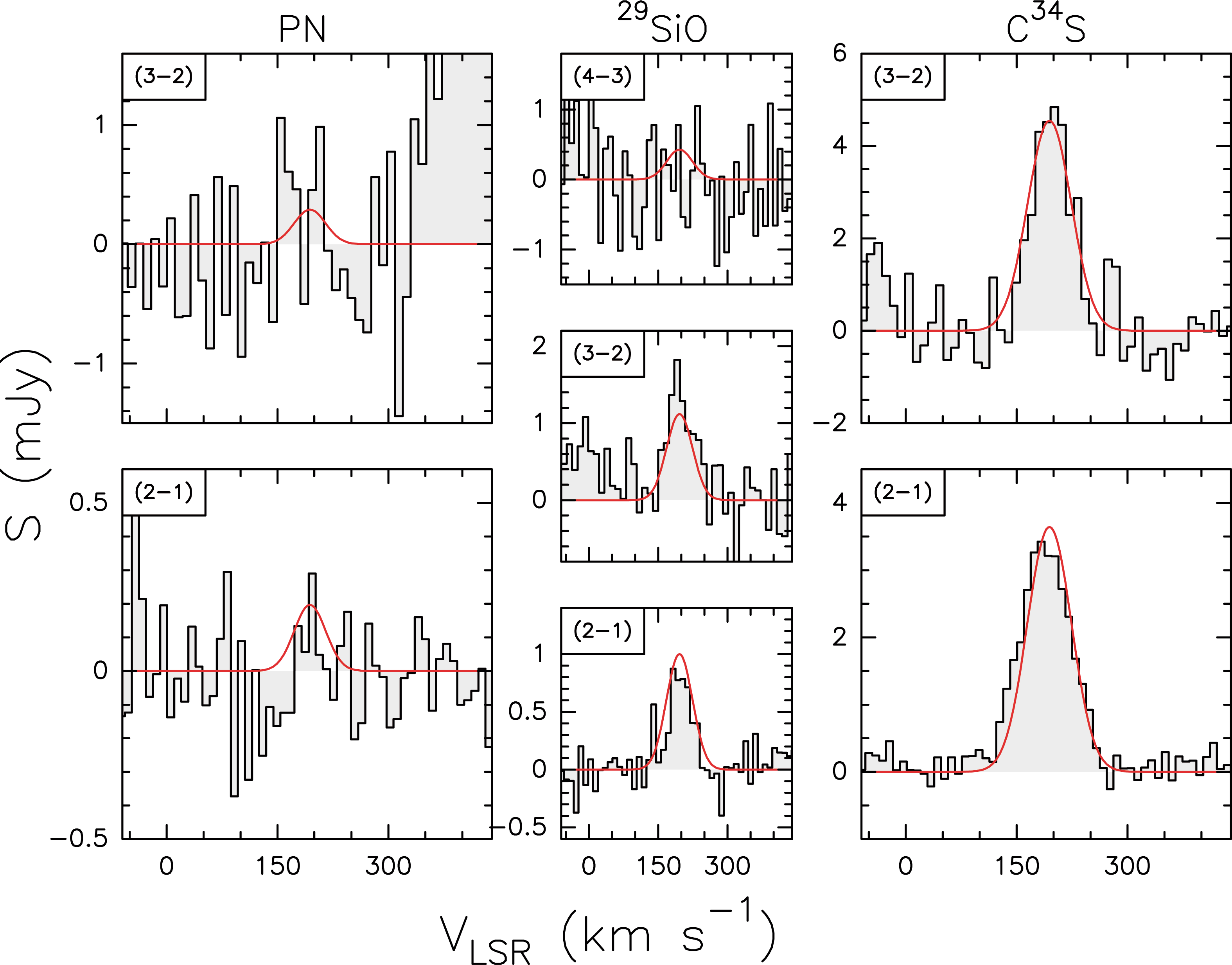}
  \caption{Spectra of PN, $^{29}$SiO, and C$^{34}$S towards GMC 8 are shown with grey histograms. The transition is indicated in the upper left of each panel. The red lines indicate the LTE best fits except for PN, for which the red line indicates an upper limit.}
     \label{figure:cloud8}
\end{figure*}

\begin{figure*}
\centering
\includegraphics[width=14cm]{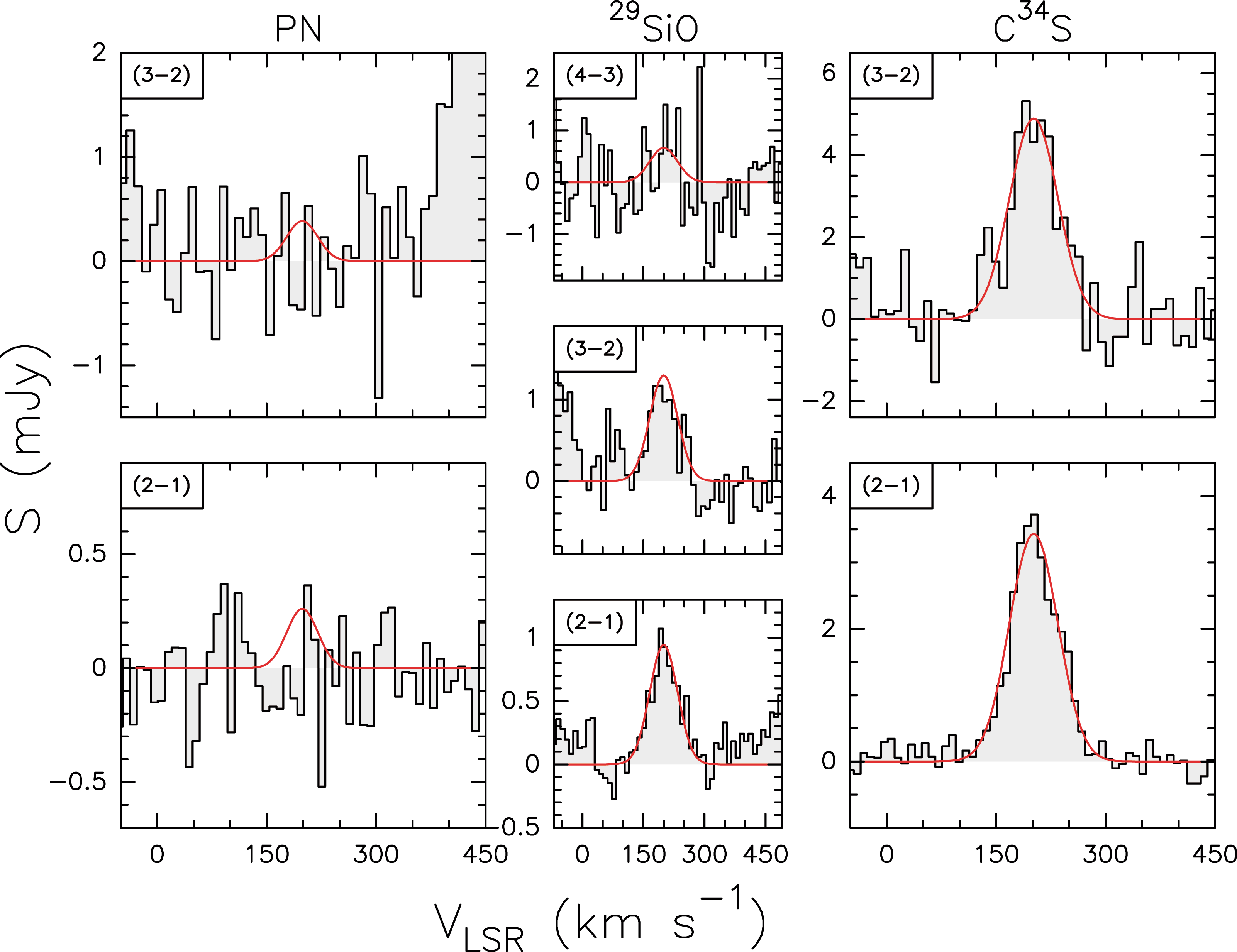}
  \caption{Spectra of PN, $^{29}$SiO, and C$^{34}$S towards GMC 9 are shown with grey histograms. The transition is indicated in the upper left of each panel. The red lines indicate the LTE best fits, except for PN, for which the red line indicates an upper limit.}
     \label{figure:cloud9}
\end{figure*}

\begin{figure*}
\centering
\includegraphics[width=14cm]{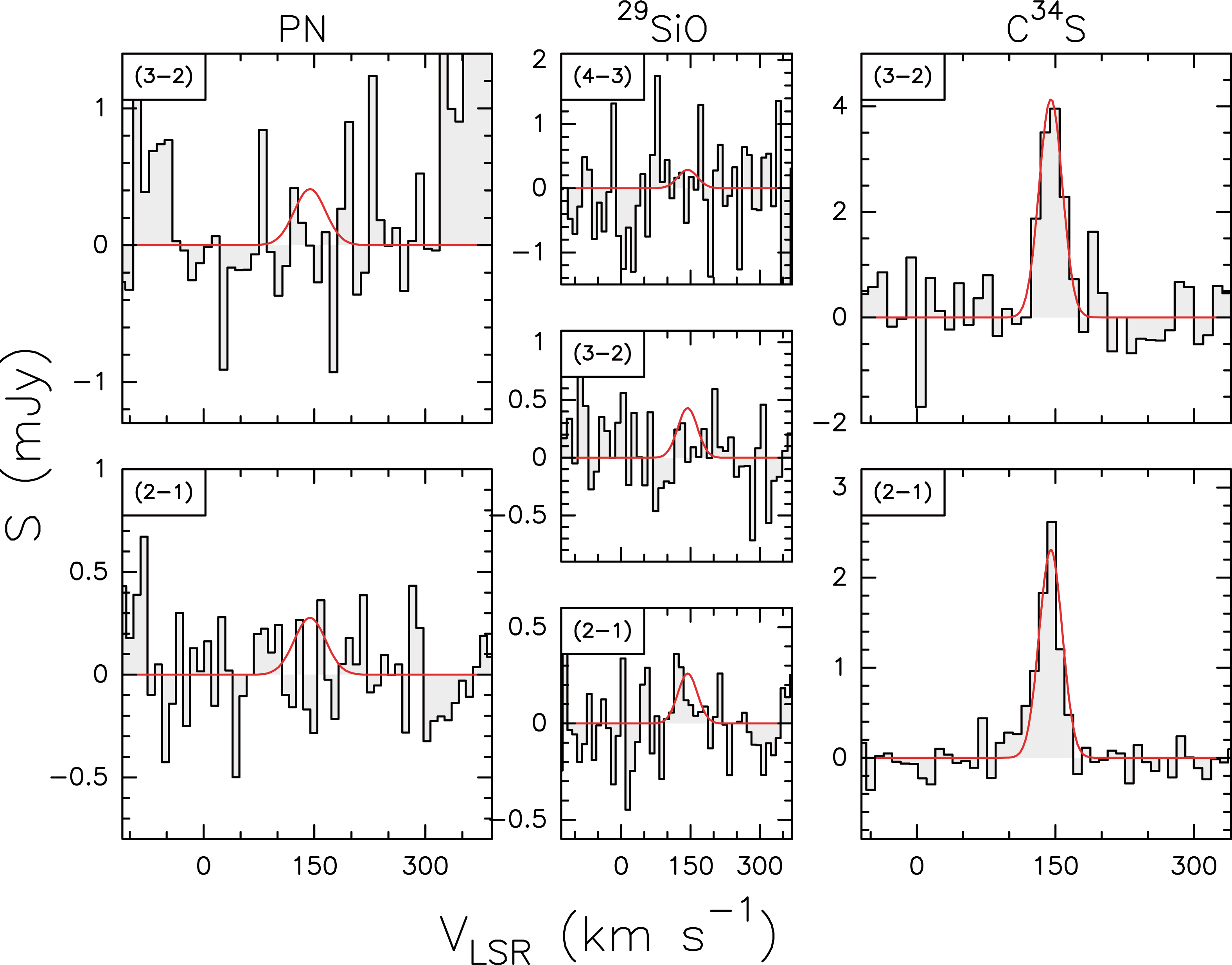}
  \caption{Spectra of PN, $^{29}$SiO, and C$^{34}$S towards GMC 10 are shown with grey histograms. The transition is indicated in the upper left of each panel. The red lines indicate the LTE best fit for C$^{34}$S, while for $^{29}$SiO and PN the red line indicates an upper limit.}
     \label{figure:cloud10}
\end{figure*}

Figures \ref{figure:cloud4} and \ref{figure:cloud6} show the spectra and the LTE fits of $^{29}$SiO and C$^{34}$S analysed towards the GMCs 4 and 6, respectively.
Figures \ref{figure:cloud1} to \ref{figure:cloud10} show the spectra, LTE fits, and upper limits of PN, $^{29}$SiO, and C$^{34}$S analysed towards the rest of the GMCs. 
Table \ref{table:appendix_clouds} lists the physical parameters of the emission of PN, $^{29}$SiO, and C$^{34}$S towards the GMCs with no PN detection, obtained from the LTE analysis.

\begin{table*}
\begin{center}
\begin{threeparttable}
\caption{Physical parameters of the emission of PN, $^{29}$SiO, and C$^{34}$S towards the GMCs 1, 2, 3, 7, 8, 9, and 10 obtained from the LTE analysis. }
\begin{tabular}{ c c c c c } 
\hline
  \noalign{\vskip 3pt}    
Molecule & \textit{N} (cm$^{-2}$) $\times$10$^{13}$ & \textit{T$_{\rm ex}$} (K) & v$_{\rm LSR}$ (km s$^{-1}$) & FWHM (km s$^{-1}$) \\ [2pt]
  \hline
  \noalign{\vskip 4pt}
  \multicolumn{5}{c}{GMC 1} \\ [2pt]
  \hline
  \noalign{\vskip 4pt}
  PN & $<$0.24 & 4.4\tnote{a} & 315\tnote{a} & 50\tnote{a} \\
  $^{29}$SiO & 2.0$\pm$0.5 & 4.9$\pm$0.4 & 313$\pm$3 & 68$\pm$8 \\
  C$^{34}$S & 9.2$\pm$1.5 & 5.0$\pm$0.4 & 315.4$\pm$1.7 & 68$\pm$4 \\ [2pt]
  \hline
  \noalign{\vskip 4pt}
  \multicolumn{5}{c}{GMC 2} \\ [2pt]
  \hline
  \noalign{\vskip 4pt}
  PN & $<$0.21 & 4.4\tnote{a} & 328\tnote{a} & 50\tnote{a} \\
  $^{29}$SiO & 0.56$\pm$0.11 & 5\tnote{a} & 329$\pm$4 & 38$\pm$9 \\
  C$^{34}$S & 7$\pm$3 & 4.4$\pm$0.6 & 328.2$\pm$2.2 & 44$\pm$5 \\ [2pt]
  \hline
  \noalign{\vskip 4pt}
  \multicolumn{5}{c}{GMC 3} \\ [2pt]
  \hline
  \noalign{\vskip 4pt}
  PN & $<$0.4 & 4.4\tnote{a} & 283\tnote{a} & 50\tnote{a} \\
  %SiO & 22.1$\pm$1.9 & 5.63$\pm$0.18 & 279.9$\pm$0.7 & 58.4$\pm$1.7 \\ 
  $^{29}$SiO & 4.3$\pm$0.6 & 5.5$\pm$0.3 & 283.2$\pm$1.9 & 60$\pm$4 \\
  C$^{34}$S & 32.8$\pm$1.3 & 6.11$\pm$0.16 & 283.6$\pm$0.4 & 60.3$\pm$1.0 \\ [2pt]
  \hline
  \noalign{\vskip 4pt}
  \multicolumn{5}{c}{GMC 7} \\ [2pt]
  \hline
  \noalign{\vskip 4pt}
  PN & $<$0.3 & 4.4\tnote{a} & 184\tnote{a} & 50\tnote{a} \\
  %SiO\tnote{c} & 22.4$\pm$2.2 & 5.07$\pm$0.15 & 183.0\tnote{a} & 58.3\tnote{a} \\
  & 5.8$\pm$0.9 & 5.6$\pm$0.4 & 125.2\tnote{a} & 60.2\tnote{a} \\
  $^{29}$SiO\tnote{b} & 3.0$\pm$0.4 & 5.0$\pm$0.3 & 183.0\tnote{a} & 58.3\tnote{a} \\
  & 1.5$\pm$0.5 & 4.7$\pm$0.6 & 125.2\tnote{a} & 60.2\tnote{a} \\
  C$^{34}$S\tnote{b} & 27.4$\pm$2.3 & 5.08$\pm$0.08 & 183.0$\pm$1.6 & 58.3$\pm$2.0 \\
  & 11.5$\pm$1.9 & 4.88$\pm$0.17 & 125.2$\pm$4.1 & 60.2$\pm$5.3 \\ [2pt]
  \hline
  \noalign{\vskip 4pt}
  \multicolumn{5}{c}{GMC 8} \\ [2pt]
  \hline
  \noalign{\vskip 4pt}
  PN & $<$0.19 & 4.4\tnote{a} & 195\tnote{a} & 50\tnote{a} \\
  $^{29}$SiO & 2.5$\pm$1.7 & 3.6$\pm$0.5 & 197$\pm$6 & 65$\pm$13 \\
  C$^{34}$S & 7$\pm$3 & 4.4$\pm$0.6 & 194.6$\pm$1.7 & 68$\pm$4 \\ [2pt]
  \hline
  \noalign{\vskip 4pt}
  \multicolumn{5}{c}{GMC 9} \\ [2pt]
  \hline
  \noalign{\vskip 4pt}
  PN & $<$0.3 & 4.4\tnote{a} & 200\tnote{a} & 50\tnote{a} \\
  $^{29}$SiO & 1.9$\pm$0.8 & 4.2$\pm$0.6 & 200$\pm$7 & 80$\pm$15 \\
  C$^{34}$S & 9.7$\pm$1.5 & 4.9$\pm$0.3 & 201.5$\pm$1.8 & 77$\pm$4 \\ [2pt]
  \hline
  \noalign{\vskip 4pt}
  \multicolumn{5}{c}{GMC 10} \\ [2pt]
  \hline
  \noalign{\vskip 4pt}
  PN & $<$0.3 & 4.4\tnote{a} & 145\tnote{a} & 50\tnote{a} \\
  $^{29}$SiO & $<$0.3 & 5\tnote{a} & 145\tnote{a} & 50\tnote{a} \\
  C$^{34}$S & 2.0$\pm$0.5 & 6.2$\pm$0.9 & 145.1$\pm$1.1 & 29$\pm$3 \\ [2pt]
  \hline
  \noalign{\vskip 5pt}
\end{tabular}
\begin{tablenotes}
    \item[a] Parameter fixed.
    \item[b] Two velocity components were used in the analysis.
  \end{tablenotes}
\label{table:appendix_clouds}
\end{threeparttable}
\end{center}
\end{table*}

\end{appendix}

\end{document}